\documentclass[12pt,a4paper]{article}

\usepackage{fixltx2e}
\usepackage{amsfonts} 

\usepackage[font=small]{caption}
\usepackage[text={450pt,650pt},headheight={14.5pt},centering]{geometry}

\usepackage{lmodern}
\usepackage{graphicx}
\usepackage{subfig}
\usepackage{booktabs}
\usepackage{multirow}
\usepackage{dsfont}
\usepackage{enumitem}
\usepackage{cite}
\usepackage{collref}

\usepackage{tikz}
\usetikzlibrary{plotmarks,calc,decorations.markings,decorations.pathmorphing,decorations.pathreplacing,patterns,matrix}

\tikzstyle arrow=[thick,rounded corners=18pt,-latex]
\tikzstyle box=[draw,rounded corners,outer sep=4pt]
\tikzstyle B node=[outer sep=0pt]
\tikzstyle Q node=[inner sep=1pt,outer sep=0pt]

\tikzset{%
  scalar/.style={thick,double},
  fermion/.style={thick,double,dashed},
  gluon/.style={thick,double,decorate,decoration={snake,amplitude=.25mm,segment length=1.25mm,post length=0.2mm,pre length=0.2mm}},
  hyp scalar/.style={thick},
  hyp fermion/.style={thick,dashed}
}

\usepackage[intlimits]{amsmath}
\usepackage{amssymb}
\usepackage{mathtools}
\usepackage{slashed}
\usepackage{braket}

\usepackage{hyperref}

\usepackage{xspace}

\usepackage{accents}

\numberwithin{equation}{section}

\makeatletter
\let\old@startsection=\@startsection
\let\oldl@section=\l@section
\renewcommand{\@startsection}[6]{\old@startsection{#1}{#2}{#3}{#4}{#5}{#6\mathversion{bold}}}
\renewcommand{\l@section}[2]{\oldl@section{\mathversion{bold}#1}{#2}}
\makeatother

\def\XXint#1#2#3{{\setbox0=\hbox{$#1{#2#3}{\int}$}
    \vcenter{\hbox{$#2#3$}}\kern-.5\wd0}}

\newcommand{\AdS}{\text{AdS}}
\newcommand{\CFT}{\text{CFT}}
\newcommand{\Sphere}{\mathrm{S}}
\newcommand{\Torus}{\mathrm{T}}

\newcommand{\alg}[1]{\mathrm{#1}}
\newcommand{\grp}[1]{\mathrm{#1}}

\newcommand{\algSL}{\alg{sl}}
\newcommand{\grpSL}{\grp{SL}}

\newcommand{\algSU}{\alg{su}}

\newcommand{\algU}{\alg{u}}

\newcommand{\algPSU}{\alg{psu}}

\newcommand{\gen}[1]{\mathbf{#1}}

\newcommand{\Integers}{\mathbb{Z}}

\newcommand{\superN}{\mathcal{N}}

\newcommand{\ie}{\textit{i.e.}\xspace}
\newcommand{\eg}{\textit{e.g.}\xspace}

\renewcommand{\Re}{\mathop{\mathrm{Re}}}
\renewcommand{\Im}{\mathop{\mathrm{Im}}}

\newcommand{\sL}{\mbox{\tiny L}}
\newcommand{\sR}{\mbox{\tiny R}}

\newcommand{\BES}{\text{BES}}

\newcommand{\AFS}{\text{AFS}}
\newcommand{\HL}{\text{HL}}

\newcommand{\curvearrowurl}{\curvearrowleft}
\newcommand{\curvearrowdlr}{\rotatebox[origin=c]{180}{$\curvearrowleft$}}
\newcommand{\inturl}{\;{\int \negthickspace \negthickspace \negthickspace\negthickspace \negthinspace \curvearrowurl}\mbox{ }\,} 
\newcommand{\intdlr}{\;{\int \negthickspace \negthickspace \negthickspace \negthinspace \curvearrowdlr}\mbox{ }\,}

\begin{document}
\thispagestyle{empty}

% \begin{flushright}\footnotesize\ttfamily
% \end{flushright}
\vspace*{2em}

\begin{center}
  \textbf{\Large\mathversion{bold} 
  On the singularities of \\ the R-R $\AdS_3 \times \Sphere^3 \times \Torus^4$ S matrix}

  \vspace{2em}

  \textrm{\large 
    Olof Ohlsson Sax${}^1$ 
    and Bogdan Stefa\'nski, jr.${}^{2,3}$ 
  } 

  \vspace{4em}

  \begingroup\itshape
  ${}^1$Nordita, Stockholm University and KTH Royal Institute of Technology,\\
  Roslagstullsbacken 23, SE-106 91 Stockholm, Sweden\\[0.2cm]
  ${}^2$Centre for Mathematical Science, City, University of London,\\ Northampton Square, EC1V 0HB, London, UK\\[0.2cm]
  ${}^{3}$ Perimeter Institute for Theoretical Physics
\\
 31 Caroline Street N, Waterloo, ON N2L 2Y5, Canada\footnote{Address until July 2019.}
  \par\endgroup

  \vspace{1em}

  \texttt{olof.ohlsson.sax@nordita.org, \\ Bogdan.Stefanski.1@city.ac.uk}

  %%%%%%%% 

\end{center}

\vspace{3em}

\begin{abstract}\noindent
We investigate the analytic properties of the exact magnon S matrix of string theory on 
$\AdS_3\times\Sphere^3\times\Torus^4$ with R-R flux. We show that the previously proposed dressing factors have the exact double-pole/zero structure  expected from Landau box diagrams. 
This constitutes a strong consistency check of our dressing factors, much as the Dorey-Hofman-Maldacena poles do for the all-loop dressing factor in $\AdS_5\times\Sphere^5$.
\end{abstract}

%%%%%%%%%%%%%%%%%%%%%%%%%%%%%%%%%%%%%%%%%%%%%%%%%%%%%%%%%%%%%%%%%%%%%%%%%%% 
\newpage

\tableofcontents

\section{Introduction}
\label{sec:introduction}

Determining the string theory spectrum in general spacetime backgrounds can be a formidable challenge. In some cases, however, the worldsheet theory may have special properties that can be used to solve the spectral problem. For example, the theory might be a WZW conformal field theory~\cite{Maldacena:2001km} or , as in the case of plane-wave backgrounds, the gauge-fixed worldsheet theory may be free~\cite{Metsaev:2001bj,Berenstein:2002jq,Metsaev:2002re}. More recently, certain $\AdS_{d+1}\times {\cal M}_{9-d}$ backgrounds have been shown to be integrable non-relativistic 2d worldsheet theories.\footnote{See~\cite{Arutyunov:2009ga,Beisert:2010jr} and references therein.} Such cases are  physically interesting, since they are holographically dual to $d$-dimensional gauge theories~\cite{Maldacena:1997re}, giving an exact tool for strong coupling computations.

In these integrable holographic backgrounds, the exact worldsheet S matrix is strongly constrained by symmetries, leaving only a small number of scalar \textit{dressing factors} unfixed. These dressing factors in turn satisfy crossing equations~\cite{Janik:2006dc}. Just as in relativistic integrable field theories, given a particular solution of the crossing equation, new solutions can be found by multiplying the S matrix by solutions of the homogeneous crossing equation known as Castillejo-Dalitz-Dyson (CDD) factors~\cite{Castillejo:1955ed}. Extra input, such as the presence of poles in the S matrix is required to determine the physically-correct CDD factor. In the case of $\AdS_5\times \Sphere^5$, the S matrix has simple poles associated to the presence of bound states~\cite{Dorey:2006dq} and Dorey-Hofman-Maldacena (DHM) double poles due to the exchanges of pairs of such bound states~\cite{Dorey:2007xn}. It is believed that these fix the dressing factor to precisely the Beisert-Hern\'andez-L\'opez/Beisert-Eden-Staudacher (BES) one~\cite{Beisert:2006ib,Beisert:2006ez}.

Strings on $\AdS_3\times\Sphere^3\times\Torus^4$ are also believed to be an integrable theory~\cite{Babichenko:2009dk,Cagnazzo:2012se}. In this case, symmetries fix the exact worldsheet S matrix up to \textit{four} dressing factors~\cite{Borsato:2014exa,Borsato:2014hja,Lloyd:2014bsa}. When the background is supported by R-R flux only, solution of the corresponding crossing equations were found in~\cite{Borsato:2013hoa,Borsato:2016kbm,Borsato:2016xns} and shown to have the expected simple poles associated with $\AdS_3$ bound states. In this paper we analyse the structure of Landau diagrams for the exchanges of pairs of bound states, generalising the work of~\cite{Dorey:2007xn} to $\AdS_3$ and determine the expected location of DHM double poles. We then show that the dressing factors found in~\cite{Borsato:2013hoa,Borsato:2016kbm,Borsato:2016xns}  have precisely such double poles providing strong evidence for the validity of these solutions.

In section~\ref{sec:expected-poles} we discuss the set of simple and double poles we expect the S matrix to have based on Landau diagrams. We begin in section~\ref{sec:DHM-review} with a review of the $\AdS_5$ analysis of~\cite{Dorey:2007xn} which we extend to $\AdS_3$ in section~\ref{sec:DHM-AdS3}. In section~\ref{sec:AdS3-comparison} we show that the S matrix and dressing phases of~\cite{Borsato:2013hoa,Borsato:2016kbm,Borsato:2016xns} have exactly the expected set of simple and double poles. Our conclusions are given in section~\ref{sec:conclusions}. In appendix~\ref{sec:naturalness} we discuss a particularly simple solution of the crossing equations which does not have DHM double poles. This solution illustrates the importance of DHM poles in identifying the physically relevant dressing factor. In appendix~\ref{sec:matrix-representations}, we present explicit bound state representations and, in appendix~\ref{sec:spin-chain-bound-states}, we review the relation between bound states and poles of the S matrix.

\section{Expected poles of the S matrix}
\label{sec:expected-poles}

\begin{figure}
  \centering

  \subfloat[S channel\label{fig:S-channel-generic}]{
    \begin{tikzpicture}
      \useasboundingbox (-2,-2.5) rectangle (2,2.5);

      \coordinate (i1) at (-1,-2);
      \coordinate (i2) at (+1,-2);

      \coordinate (o1) at (-1,+2);
      \coordinate (o2) at (+1,+2);

      \coordinate (v1) at (0,-1);
      \coordinate (v2) at (0,+1);

      \draw [thick] (i1) -- (v1);
      \draw [thick] (i2) -- (v1);

      \draw [thick,double] (v1) -- (v2);

      \draw [thick] (v2) -- (o1);
      \draw [thick] (v2) -- (o2);

      \fill (v1) circle [radius=0.1];
      \fill (v2) circle [radius=0.1];

      \node at (i1) [anchor=east] {\small $p_1$};
      \node at (i2) [anchor=west] {\small $p_2$};

      \node at (o1) [anchor=east] {\small $p_2$};
      \node at (o2) [anchor=west] {\small $p_1$};

      \node at ($0.5*(v1)+0.5*(v2)$) [rotate=90,anchor=north] {\small $p_1+p_2$};

    \end{tikzpicture}
  }
  \hspace{1cm}
  \subfloat[T channel\label{fig:T-channel-generic}]{
    \begin{tikzpicture}
      \useasboundingbox (-2,-2.5) rectangle (2,2.5);

      \coordinate (i1) at (-1,-2);
      \coordinate (i2) at (+1,-2);

      \coordinate (o1) at (-1,+2);
      \coordinate (o2) at (+1,+2);

      \coordinate(v1) at (-1,-0.5);
      \coordinate (v2) at (+1,+0.5);

      \draw [thick] (i1) -- (v1);
      \draw [thick] (i2) -- (v2);

      \draw [thick,double] (v1) -- (v2);

      \draw [thick] (v1) -- (o1);
      \draw [thick] (v2) -- (o2);

      \fill (v1) circle [radius=0.1];
      \fill (v2) circle [radius=0.1];

      \node at (i1) [anchor=east] {\small $p_1$};
      \node at (i2) [anchor=west] {\small $p_2$};

      \node at (o1) [anchor=east] {\small $p_2$};
      \node at (o2) [anchor=west] {\small $p_1$};

      \node at ($0.5*(v1)+0.5*(v2)$) [rotate=27,anchor=north] {\small $p_1-p_2$};
    \end{tikzpicture}
  }
  \caption{S and T channel exchange of bound states. The external lines correspond to fundamental excitations and the double line to an exchanged on-shell bound state.\label{fig:S-and-T-channel-generic}}
\end{figure}
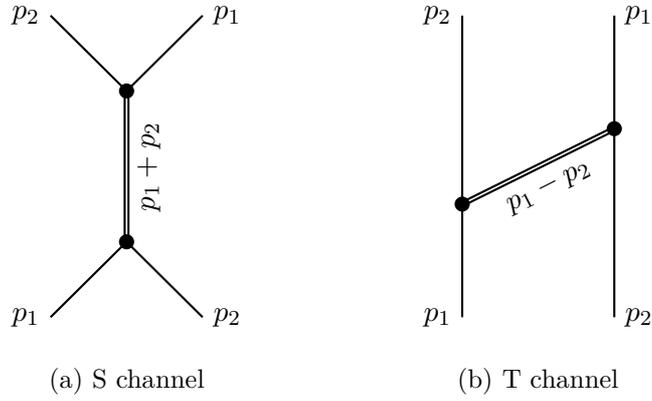

\begin{figure}
  \centering

  \begin{tikzpicture}
    \coordinate (i1) at (-2.75,-2.5);
    \coordinate (i2) at (+2,-2.5);

    \coordinate (o1) at (-2.75,+2.5);
    \coordinate (o2) at (+2,+2.5);

    \coordinate (v1) at (-1,-0.75);
    \coordinate (v2) at (+1,-1.5);
    \coordinate (v3) at (+1,+1.5);
    \coordinate (v4) at (-1,+0.75);

    \draw [thick] (i1) -- (v1);
    \draw [thick] (i2) -- (v2);
    \draw [thick] (v3) -- (o2);
    \draw [thick] (v4) -- (o1);

    \draw [thick, double] (v1) -- (v2) -- (v3) -- (v4) -- (v1);

    \fill (v1) circle [radius=0.1];
    \fill (v2) circle [radius=0.1];
    \fill (v3) circle [radius=0.1];
    \fill (v4) circle [radius=0.1];
  \end{tikzpicture}
  
  \caption{Typical box diagram giving rise to a double pole in the S matrix.\label{fig:box-generic}}
\end{figure}
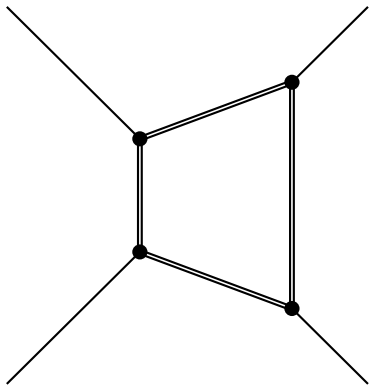

Poles in the S matrix are intimately related to bound states in the spectrum.\footnote{See appendix~\ref{sec:spin-chain-bound-states} for a discussion of the relation between singularities of the S matrix and bound states.} A simple pole can be represented by a diagram where two excitations scatter through the exchange of a single on-shell bound state. There are two basic topologies, referred to as the S and T channels, as shown in figure~\ref{fig:S-and-T-channel-generic}.
In order to obtain a double pole, we need a scattering diagram where two exchanged bound states simultaneously go on shell. A typical such diagram is shown in figure~\ref{fig:box-generic}.

\subsection{Bound states and S matrix poles in $\AdS_5 \times \Sphere^5$}
\label{sec:DHM-review}

In this sub-section we briefly review the bound states and Landau diagrams leading to poles in the $\AdS_5 \times \Sphere^5$ S matrix following closely the discussion in~\cite{Dorey:2007xn}. World-sheet excitations in $\AdS_5 \times \Sphere^5$ transform in short representations of the centrally extended $\algPSU(2|2)^2$ algebra preserved by the light-cone gauge Hamiltonian. They are conveniently parametrised by spectral parameters $x^{\pm}$ which are related to the energy and momentum by
\begin{equation}\label{eq:p-and-E-of-xpm}
  e^{ip} = \frac{x^+}{x^-} , \qquad
  E = -\frac{ih}{2} \biggl( x^+ - \frac{1}{x^+} - x^- + \frac{1}{x^-} \biggr) ,
\end{equation}
where $h$ is the integrable coupling constant. The fact that the representation is short is encoded in the additional constraint
\begin{equation}\label{eq:shortening-condition-xpm}
  x^+ + \frac{1}{x^+} - x^- - \frac{1}{x^-} = \frac{2iM}{h} ,
\end{equation}
where $M\in\mathbb{Z}^+$ is the bound state number. A fundamental representation has $M=1$ while a bound state has $M > 1$. The corresponding $\algPSU(2|2)$ representation has a highest weight state transforming as $(M+1,1)$ under the bosonic $\algSU(2) \times \algSU(2)$ subalgebra. A physical excitation in $\AdS_5 \times \Sphere^5$ has spectral parameters satisfying $|x^{\pm}| > 1$. The shortening condition~\eqref{eq:shortening-condition-xpm} together with~\eqref{eq:p-and-E-of-xpm} lead to the familiar dispersion relation
\begin{equation}\label{eq:dispersion-relation}
  E = \sqrt{M^2 + 4h^2\sin^2\frac{p}{2}} .
\end{equation}

\begin{figure}
  \centering
  \subfloat[\label{fig:vertex-FF-B}]{
    \begin{tikzpicture}
      \useasboundingbox (-2,-1.75) rectangle (2,1.5);

      \coordinate (i1) at (-1,-1);
      \coordinate (i2) at (+1,-1);

      \coordinate (v) at (0,0);

      \coordinate (o) at (0,+1);

      \draw [thick] (i1) -- (v);
      \draw [thick] (i2) -- (v);

      \draw [thick,double] (v) -- (o);

      \fill (v) circle [radius=0.1];

      \node at (i1) [anchor=east] {\small $x^{\pm}$};
      \node at (i2) [anchor=west] {\small $y^{\pm}$};
      \node at (o) [anchor=south] {\small $z^{\pm}$};
    \end{tikzpicture}
  }
  \hspace{1cm}
  \subfloat[\label{fig:vertex-B-FF}]{
    \begin{tikzpicture}
      \useasboundingbox (-2,-1.75) rectangle (2,1.5);

      \coordinate (o1) at (-1,+1);
      \coordinate (o2) at (+1,+1);

      \coordinate (v) at (0,0);

      \coordinate (i) at (0,-1);

      \draw [thick] (o1) -- (v);
      \draw [thick] (o2) -- (v);

      \draw [thick,double] (v) -- (i);

      \fill (v) circle [radius=0.1];

      \node at (o1) [anchor=east] {\small $y^{\pm}$};
      \node at (o2) [anchor=west] {\small $x^{\pm}$};
      \node at (i) [anchor=north] {\small $z^{\pm}$};
    \end{tikzpicture}
  }
  \caption{Basic vertices\label{fig:vertices}}
\end{figure}
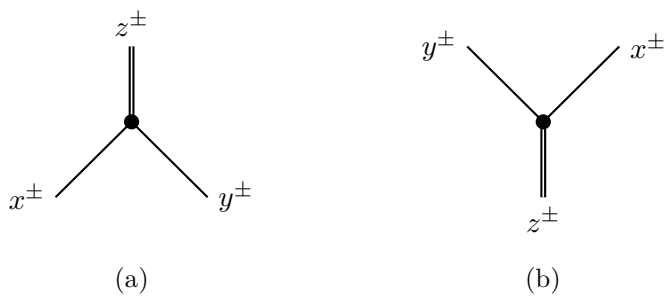

The on-shell diagrams discussed above can be decomposed into trivalent vertices of the type shown in figure~\ref{fig:vertices}, where we have introduced the spectral parameters $x^{\pm}$, $y^{\pm}$ and $z^{\pm}$ corresponding to the $\algPSU(2|2)$ representations at each leg, with
the three excitations have bound state numbers $M_x$, $M_y$ and $M_z$, respectively.
Imposing energy and momentum conservation at the vertex in figure~\ref{fig:vertex-FF-B} leads to the equations\footnote{The trivalent vertex shown in figure~\ref{fig:vertex-B-FF} is related to the one in figure~\ref{fig:vertex-FF-B} by time reversal and leads to exactly the same set of equations.}
\begin{equation}
  \begin{gathered}
    \biggl( x^+ - \frac{1}{x^+} - x^- + \frac{1}{x^-} \biggr)
    +
    \biggl( y^+ - \frac{1}{y^+} - y^- + \frac{1}{y^-} \biggr)
    =
    \biggl( z^+ - \frac{1}{z^+} - z^- + \frac{1}{z^-} \biggr) , \\
    \frac{x^+}{x^-} \frac{y^+}{y^-} = \frac{z^+}{z^-} .
  \end{gathered}
\end{equation}
These equations have a number of simple solutions. If we assume that all three involved excitations are in the physical region we find
\begin{equation}\label{eq:bound-state-psu22-su2}
  x^- = y^+ , \qquad
  z^+ = x^+ , \qquad
  z^- = y^- ,
\end{equation}
or
\begin{equation}\label{eq:bound-state-psu22-sl2}
  x^+ = y^- , \qquad
  z^+ = y^+ , \qquad
  z^- = x^- .
\end{equation}
From either solution we find the relation
\begin{equation}
  \biggl( x^+ + \frac{1}{x^+} - x^- - \frac{1}{x^-} \biggr)
  +
  \biggl( y^+ + \frac{1}{y^+} - y^- - \frac{1}{y^-} \biggr)
  =
  \biggl( z^+ + \frac{1}{z^+} - z^- - \frac{1}{z^-} \biggr) ,
\end{equation}
or
\begin{equation}
  M_z = M_x + M_y .
\end{equation}
Hence, in these cases the vertex has a simple interpretation in terms of two physical excitations fusing into a larger physical bound state. When the two incoming excitations are fundamental ($M_x = M_y = 1$), the solutions~\eqref{eq:bound-state-psu22-su2} and~\eqref{eq:bound-state-psu22-sl2} correspond to the so called $\algSU(2)$ and $\algSL(2)$ bound states, respectively. As we will see below, only the $\algSU(2)$ bound state appears in the physical spectrum.

Let us now consider the case where the spectral parameters of one of the incoming excitations are outside the physical region. When $|x^{\pm}| \,,\,|z^{\pm}|> 1$, $|y^{\pm}| < 1$ we find solutions
\begin{equation}\label{eq:bound-state-psu22-su2-crossed}
  x^- = \frac{1}{y^-} , \qquad
  z^+ = x^+ , \qquad
  z^- = \frac{1}{y^+} ,
\end{equation}
and
\begin{equation}\label{eq:bound-state-psu22-sl2-crossed}
  x^+ = \frac{1}{y^+} , \qquad
  z^+ = \frac{1}{y^-} , \qquad
  z^- = x^- .
\end{equation}
These spectral parameters satisfy
\begin{equation}
  \biggl( x^+ + \frac{1}{x^+} - x^- - \frac{1}{x^-} \biggr)
  -
  \biggl( y^+ + \frac{1}{y^+} - y^- - \frac{1}{y^-} \biggr)
  =
  \biggl( z^+ + \frac{1}{z^+} - z^- - \frac{1}{z^-} \biggr) ,
\end{equation}
or
\begin{equation}
  M_z = M_x - M_y .
\end{equation}
Note that this process is possible only if $M_x > M_y$. In that case the vertex corresponds to a process where a bound state fusing with an antiparticle to produce a new state with a lower bound state number.

Starting with the solutions~\eqref{eq:bound-state-psu22-su2} or~\eqref{eq:bound-state-psu22-sl2} we can obtain~\eqref{eq:bound-state-psu22-su2-crossed} or~\eqref{eq:bound-state-psu22-sl2-crossed}, respectively, by the transformation
\begin{equation}
  x^{\pm} \to z^{\pm} , \qquad
  y^{\pm} \to \frac{1}{x^{\pm}} , \qquad
  z^{\pm} \to y^{\pm} ,
\end{equation}
which can be interpreted as relabelling followed by crossing one of the legs by sending $y^{\pm} \to 1/y^{\pm}$.

If instead we have $|x^{\pm}| < 1$, $|y^{\pm}| > 1$, and $|z^{\pm}| > 1$ we find the same set of solutions as above, but with $x^{\pm}$ and $y^{\pm}$ swapped. Finally, there are solutions where two or more of the spectral parameters are inside the unit circle. These solutions are again related to the ones discussed above by simple transformations.

\subsubsection{Simple poles}

Let us now combine the above vertices to form physical scattering diagrams, starting with the diagrams in figure~\ref{fig:S-and-T-channel-generic}, which lead to simple poles in the S matrix.

\paragraph{S channel pole.}

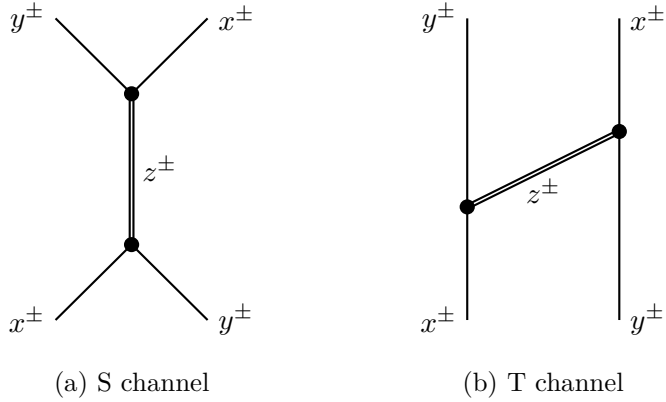
\begin{figure}
  \centering

  \subfloat[S channel\label{fig:S-channel-x}]{
    \begin{tikzpicture}
      \useasboundingbox (-2,-2.5) rectangle (2,2.5);

      \coordinate (i1) at (-1,-2);
      \coordinate (i2) at (+1,-2);

      \coordinate (o1) at (-1,+2);
      \coordinate (o2) at (+1,+2);

      \coordinate (v1) at (0,-1);
      \coordinate (v2) at (0,+1);

      \draw [thick] (i1) -- (v1);
      \draw [thick] (i2) -- (v1);

      \draw [thick,double] (v1) -- (v2);

      \draw [thick] (v2) -- (o1);
      \draw [thick] (v2) -- (o2);

      \fill (v1) circle [radius=0.1];
      \fill (v2) circle [radius=0.1];

      \node at (i1) [anchor=east] {\small $x^{\pm}$};
      \node at (i2) [anchor=west] {\small $y^{\pm}$};

      \node at (o1) [anchor=east] {\small $y^{\pm}$};
      \node at (o2) [anchor=west] {\small $x^{\pm}$};

      \node at ($0.5*(v1)+0.5*(v2)$) [anchor=west] {\small $z^{\pm}$};

    \end{tikzpicture}
  }
  \hspace{1cm}
  \subfloat[T channel\label{fig:T-channel-x}]{
    \begin{tikzpicture}
      \useasboundingbox (-2,-2.5) rectangle (2,2.5);

      \coordinate (i1) at (-1,-2);
      \coordinate (i2) at (+1,-2);

      \coordinate (o1) at (-1,+2);
      \coordinate (o2) at (+1,+2);

      \coordinate(v1) at (-1,-0.5);
      \coordinate (v2) at (+1,+0.5);

      \draw [thick] (i1) -- (v1);
      \draw [thick] (i2) -- (v2);

      \draw [thick,double] (v1) -- (v2);

      \draw [thick] (v1) -- (o1);
      \draw [thick] (v2) -- (o2);

      \fill (v1) circle [radius=0.1];
      \fill (v2) circle [radius=0.1];

      \node at (i1) [anchor=east] {\small $x^{\pm}$};
      \node at (i2) [anchor=west] {\small $y^{\pm}$};

      \node at (o1) [anchor=east] {\small $y^{\pm}$};
      \node at (o2) [anchor=west] {\small $x^{\pm}$};

      \node at ($0.5*(v1)+0.5*(v2)$) [anchor=north] {\small $z^{\pm}$};
    \end{tikzpicture}
  }
  \caption{S and T channel labelled with spectral parameters.\label{fig:S-and-T-channel-x}}
\end{figure}

We first consider the S channel diagram shown in figure~\ref{fig:S-channel-generic}, introducing spectral parameters as in figure~\ref{fig:S-channel-x}.
We focus on the case where the external excitations are fundamental, so that $M_x = M_y = 1$. We then only have a physical pole of the form\footnote{When we discuss physical scattering processes we assume that the excitations are ordered along the spatial direction so that the excitation with parameters $x^{\pm}$ is to the left of the excitation with parameters $y^{\pm}$. This means that the solutions in equation~\eqref{eq:S-pole-psu22-su2} and~\eqref{eq:S-pole-psu22-sl2} are distinct even though the algebraic expressions are related by exchanging $x^{\pm}$ and $y^{\pm}$.}
\begin{equation}\label{eq:S-pole-psu22-su2}
  x^+ = y^- , \qquad
  z^+ = y^+ , \qquad
  z^- = x^- ,
\end{equation}
or
\begin{equation}\label{eq:S-pole-psu22-sl2}
  x^- = y^+ , \qquad
  z^+ = x^+ , \qquad
  z^- = y^- ,
\end{equation}
and the exchanged bound state has $M_z = M_x + M_y = 2$. To find out which solution gives rise to a pole in the S matrix corresponding to a physical bound state, we need to examine the imaginary part of the momentum of the first particle, as discussed in appendix~\ref{sec:spin-chain-bound-states}. For~\eqref{eq:S-pole-psu22-su2} we find $\Im(-i\log x^+/x^-) > 0$ and for~\eqref{eq:S-pole-psu22-sl2} we find $\Im(-i\log x^+/x^-) < 0$, which means that the physical pole is given by~\eqref{eq:S-pole-psu22-su2}, while~\eqref{eq:S-pole-psu22-sl2} corresponds to a zero of the S matrix.\footnote{The easiest way to check the imaginary part of the momentum is to solve equation~\eqref{eq:S-pole-psu22-su2} or~\eqref{eq:S-pole-psu22-sl2} together with the shortening conditions~\eqref{eq:AdS3-shortening-condition-xpm} for $x^{\pm}$ and $y^{\pm}$ using, \eg, Mathematica, making sure to pick the branch where both $|x^{\pm}| > 1$ and $|y^{\pm}| > 1$, and then numerically evaluate $x^+/x^-$ for real values of the bound state momentum.}

\paragraph{T channel pole.}

Consider now the T channel diagram in figure~\ref{fig:T-channel-x}. If the external excitations are both fundamental ($M_x = M_y = 1$), they can not also both be in the physical region. Instead we find
\begin{equation}\label{eq:T-pole-psu22-su2}
  x^+ = \frac{1}{y^-} , \qquad
  z^+ = \frac{1}{y^+} , \qquad
  z^- = x^- , 
\end{equation}
or
\begin{equation}\label{eq:T-pole-psu22-sl2}
  x^- = \frac{1}{y^+} , \qquad
  z^+ = x^+ , \qquad
  z^- = \frac{1}{y^-} .
\end{equation}
The exchanged bound state again has $M_z = M_x + M_y = 2$. As before we check the sign of the imaginary part of the momentum of the $x$ excitation to determine when the S matrix should have a pole. For the solution~\eqref{eq:T-pole-psu22-su2} we find $\Im( -i\log x^+/x^- ) > 0$ and for~\eqref{eq:T-pole-psu22-sl2} we find $\Im( -i\log x^+/x^- ) < 0$, so that the location of the pole is given by~\eqref{eq:T-pole-psu22-su2}.

\subsubsection{Double poles}
\label{sssec:double-poles}

\begin{figure}
  \centering

  \begin{tikzpicture}
    \coordinate (i1) at (-2.75,-2.5);
    \coordinate (i2) at (+2,-2.5);

    \coordinate (o1) at (-2.75,+2.5);
    \coordinate (o2) at (+2,+2.5);

    \coordinate (v1) at (-1,-0.75);
    \coordinate (v2) at (+1,-1.5);
    \coordinate (v3) at (+1,+1.5);
    \coordinate (v4) at (-1,+0.75);

    \draw [thick] (i1) -- (v1);
    \draw [thick] (i2) -- (v2);
    \draw [thick] (v3) -- (o2);
    \draw [thick] (v4) -- (o1);

    \draw [thick, double] (v1) -- (v2) -- (v3) -- (v4) -- (v1);

    \fill (v1) circle [radius=0.1];
    \fill (v2) circle [radius=0.1];
    \fill (v3) circle [radius=0.1];
    \fill (v4) circle [radius=0.1];

    \node at (i1) [anchor=east] {\small $x_1^{\pm}$};
    \node at (i2) [anchor=west] {\small $x_2^{\pm}$};
    \node at (o1) [anchor=east] {\small $x_2^{\pm}$};
    \node at (o2) [anchor=west] {\small $x_1^{\pm}$};

    \node at ($0.5*(v1)+0.5*(v2)$) [anchor=north] {\small $z_1^{\pm}$};
    \node at ($0.5*(v2)+0.5*(v3)$) [anchor=west] {\small $y_2^{\pm}$};
    \node at ($0.5*(v3)+0.5*(v4)$) [anchor=south] {\small $z_2^{\pm}$};
    \node at ($0.5*(v4)+0.5*(v1)$) [anchor=east] {\small $y_1^{\pm}$};
  \end{tikzpicture}
  
  \caption{Box diagram with spectral parameters.\label{fig:box-x}}
\end{figure}
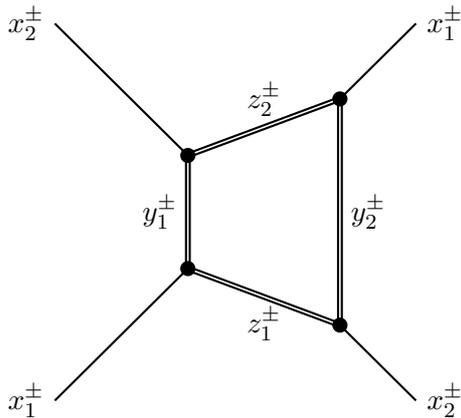

We now consider the box diagram in figure~\ref{fig:box-x}.
The two on-shell bound states $y_1^{\pm}$ and $y_2^{\pm}$ give rise to a double pole in the S matrix. The four vertices making up the box diagram are of the same type discussed above. Imposing that the double pole is located on the physical branch we find the solution\footnote{For a more detailed derivation of these conditions see~\cite{Dorey:2007xn} and the corresponding derivation for $\AdS_3 \times \Sphere^3 \times \Torus^4$ in the section~\ref{sec:DHM-AdS3}.}
\begin{equation}
  x_1^+ = \frac{1}{y_1^-} = \frac{1}{z_2^-} , \qquad
  x_1^- = y_2^- = \frac{1}{z_1^-} , \qquad
  x_2^+ = y_2^+ = \frac{1}{z_2^+} , \qquad
  x_2^- = \frac{1}{y_1^+} = \frac{1}{z_1^+} ,
\end{equation}
which leads to the relation
\begin{equation}
  \biggl( x_1^+ + \frac{1}{x_1^+} + x_1^- + \frac{1}{x_1^-} \biggr)
  -
  \biggl( x_2^+ + \frac{1}{x_2^+} + x_2^- + \frac{1}{x_2^-} \biggr)
  =
  -\frac{2i}{h} ( M_{z_1} + M_{z_2} )
%   =
%   -\frac{2i}{h} ( M_{y_1} + M_{y_2} ) .
\end{equation}
Introducing
\begin{equation}
  u_i = \frac{1}{2} \biggl( x_i^+ + \frac{1}{x_i^+} + x_i^- + \frac{1}{x_i^-} \biggr) ,
\end{equation}
we thus find double poles at the locations
\begin{equation}
  u_1 - u_2 = -\frac{2in}{h} ,
\end{equation}
where $n = (M_{z_1} + M_{z_2})/2 > 1$ is an integer.

\subsection{Bound states and S matrix poles in $\AdS_3\times \Sphere^3\times \Torus^4$}
\label{sec:DHM-AdS3}

In the case of string theory on $\AdS_3 \times \Sphere^3 \times \Torus^4$ supported by RR flux, the fundamental excitations fall into several short representations of the centrally extended $\algPSU(1|1)^4$ algebra preserved by the light-cone gauge-fixed Hamiltonian~\cite{Borsato:2013qpa,Borsato:2014exa,Borsato:2014hja}. For now we focus on the massive excitations, which fall into two representations denoted L and R. These representations are distinguished by the $\algU(1)$ charge $\gen{M}$, under which they have eigenvalue $+1$ and $-1$, respectively. In order to understand the pole structure of the S matrix it is enough to consider a centrally extended $\algPSU(1|1)^2$ sub-algebra. As in $\AdS_5 \times
\Sphere^5$, short representations can be described by spectral parameters $x^{\pm}$ which parametrise the energy and momentum as in equation~\eqref{eq:p-and-E-of-xpm} and satisfy the shortening condition
\begin{equation}\label{eq:AdS3-shortening-condition-xpm}
  x^+ + \frac{1}{x^+} - x^- - \frac{1}{x^-} = \frac{2i|M|}{h} ,
\end{equation}
where $M$ is the $\algU(1)$ charge of the representation.\footnote{%
  Note that the absolute value of $M$ appears on the right-hand side of equation~\eqref{eq:AdS3-shortening-condition-xpm} since the $\algU(1)$ eigenvalue can be both positive and negative. In $\AdS_5$ the corresponding charge is non-negative since it labels the $\algSU(2)$ charge of a highest weight state.%
}
Any short $\algPSU(1|1)^2$ representation is two-dimensional. Hence, the only difference between a fundamental excitation and a bound state in $\AdS_3$ is the $\algU(1)$ charge, which plays the role of the mass as seen in the dispersion relation~\eqref{eq:dispersion-relation}.

\begin{figure}
  \centering
  \subfloat[LLL\label{fig:AdS3-vertex-LLL}]{
      \begin{tikzpicture}
      \useasboundingbox (-2,-1.75) rectangle (2,1.75);

      \coordinate (i1) at (-1,-1);
      \coordinate (i2) at (+1,-1);

      \coordinate (v) at (0,0);

      \coordinate (o) at (0,+1.5);

      \draw [thick] (i1) -- (v);
      \draw [thick] (i2) -- (v);

      \draw [thick,double] (v) -- (o);

      \fill (v) circle [radius=0.1];

      \node at (i1) [anchor=east] {\small $x_{\sL}^{\pm}$};
      \node at (i2) [anchor=west] {\small $y_{\sL}^{\pm}$};
      \node at (o) [anchor=south] {\small $z_{\sL}^{\pm}$};

      \node at ($0.5*(i1)+0.5*(v)$) [anchor=south,rotate=45] {$\scriptscriptstyle M=1$};
      \node at ($0.5*(i2)+0.5*(v)$) [anchor=south,rotate=-45] {$\scriptscriptstyle M=m$};
      \node at ($0.5*(o)+0.5*(v)$) [anchor=south,rotate=90] {$\scriptscriptstyle M=1+m$};
    \end{tikzpicture}
  }
  \subfloat[LRR\label{fig:AdS3-vertex-LRR}]{
      \begin{tikzpicture}
      \useasboundingbox (-2,-1.75) rectangle (2,1.75);

      \coordinate (i1) at (-1,-1);
      \coordinate (i2) at (+1,-1);

      \coordinate (v) at (0,0);

      \coordinate (o) at (0,+1.5);

      \draw [thick] (i1) -- (v);
      \draw [thick] (i2) -- (v);

      \draw [thick,double] (v) -- (o);

      \fill (v) circle [radius=0.1];

      \node at (i1) [anchor=east] {\small $x_{\sL}^{\pm}$};
      \node at (i2) [anchor=west] {\small $y_{\sR}^{\pm}$};
      \node at (o) [anchor=south] {\small $z_{\sR}^{\pm}$};

      \node at ($0.5*(i1)+0.5*(v)$) [anchor=south,rotate=45] {$\scriptscriptstyle M=1$};
      \node at ($0.5*(i2)+0.5*(v)$) [anchor=south,rotate=-45] {$\scriptscriptstyle M=-m$};
      \node at ($0.5*(o)+0.5*(v)$) [anchor=south,rotate=90] {$\scriptscriptstyle M=1-m$};
    \end{tikzpicture}
  }
    
  \caption{LL and LR vertices}
\end{figure}
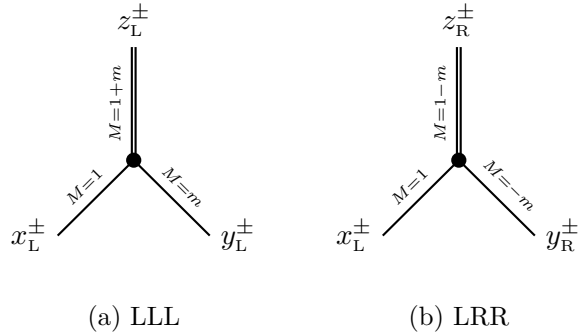

As in $\AdS_5\times\Sphere^5$, we can construct trivalent vertices from which we can build up scattering diagrams that exhibit the poles of the S matrix. Imposing conservation of energy, momentum and $\algU(1)$ charge we obtain relations between the spectral parameters of the excitations. Let us start with a fundamental L excitation with spectral parameters $x_{\sL}^{\pm}$ and charge $+1$ scattering with a state $y_{\sL}^{\pm}$ with charge $M = m \ge 1$. From charge conservation we know that we can only produce a state with charge $M = m+1$, whose spectral parameters we denote by $z_{\sL}^{\pm}$. This is illustrated in figure~\ref{fig:AdS3-vertex-LLL}.
 We find that the spectral parameters satisfy either
\begin{equation}
  x_{\sL}^- = y_{\sL}^+ , \qquad
  z_{\sL}^+ = x_{\sL}^+ , \qquad
  z_{\sL}^- = y_{\sL}^- ,
\end{equation}
or
\begin{equation}
  x_{\sL}^+ = y_{\sL}^- , \qquad
  z_{\sL}^+ = y_{\sL}^+ , \qquad
  z_{\sL}^- = x_{\sL}^- .
\end{equation}
Instead if we start with a fundamental L excitation ($x_{\sL}^{\pm}$, $+1$) and an R bound state ($y_{\sR}^{\pm}$, $-m<-1$), we can produce an R excitation with $z_{\sR}^{\pm}$ and charge $-m+1 < 0$, as illustrated in figure~\ref{fig:AdS3-vertex-LRR} with
\begin{equation}
  x_{\sL}^+ = \frac{1}{y_{\sR}^+} , \qquad
  z_{\sR}^+ = \frac{1}{x_{\sL}^-} , \qquad
  z_{\sR}^- = y_{\sR}^- ,
  \label{eq:lr-bnd1}
\end{equation}
or
\begin{equation}
  x_{\sL}^- = \frac{1}{y_{\sR}^-} , \qquad
  z_{\sR}^+ = y_{\sR}^+ , \qquad
  z_{\sR}^- = \frac{1}{x_{\sL}^+} .
  \label{eq:lr-bnd2}
\end{equation}
Using these basic vertices we can construct box diagrams similar to those of $\AdS_5 \times \Sphere^5$.

From an algebraic point of view it would appear that we can consider solutions given in equations~\eqref{eq:lr-bnd1},~\eqref{eq:lr-bnd1} also in the case $m=1$. In this case the outgoing state $z_{\sR}^{\pm}$ would have mass $1 - m = 0$. Such a hypothetical solution would correspond to a \emph{massless} bound state. String theory on $\AdS_3\times \Sphere^3\times\Torus^4$ does contain a massless multiplet. However, it is not possible to identify  the hypothetical massless bound states with the massless multiplet, for two reasons. Firstly, the massless modes arising from $\Torus^4$ form a doublet under an auxiliary $\algSU(2)_\circ$ which commutes with the centrally extended $\algPSU(1|1)^2$ algebra~\cite{Borsato:2014exa,Borsato:2014hja}. There is no such symmetry acting on the hypothetical massless bound states. Secondly, the massless bound state has opposite statistics to the physical massless excitations~\cite{Borsato:2014exa,Borsato:2014hja}. In particular, if we consider the full centrally extended $\algPSU(1|1)^4$ algebra, the highest weight state of the hypothetical massless bound state multiplet would be bosonic.\footnote{As we show in appendix~\ref{app:lr-bound-states}, the highest weight states of LR bound states have the same statistics for all values of bound state number $M$ - they are in fact bosons in the full $\algPSU(1|1)^4$ algebra.} On the other hand the physical massless representation has a fermionic highest weight state~\cite{Borsato:2014exa,Borsato:2014hja}. Because of this, we require the S matrix to not have poles that would correspond to such massless bound states.

Similarly, the representation theory allows for an $M=1$ ``bound state'' which would appear in the tensor product between a $\Torus^4$ massless mode and an $M=1$ massive excitation. This process is closely related to the one discussed in the previous paragraph, and again leads to a representation with statistics which is opposite of that of the normal fundamental massive excitations~\cite{Borsato:2013qpa}. We will therefore also require the S matrix between a massive and a massless excitation to have no corresponding poles.

Finally, we note that the representation theory does not allow for a bound state in the scattering between two massless excitations~\cite{Borsato:2016kbm,Borsato:2016xns}.

\begin{figure}
  \centering

  \subfloat[S channel\label{fig:AdS3-S-channel-LL}]{
    \begin{tikzpicture}
      \useasboundingbox (-3.5,-2.5) rectangle (3.5,2.5);

      \coordinate (i1) at (-1,-2);
      \coordinate (i2) at (+1,-2);

      \coordinate (o1) at (-1,+2);
      \coordinate (o2) at (+1,+2);

      \coordinate (v1) at (0,-1);
      \coordinate (v2) at (0,+1);

      \draw [thick] (i1) -- (v1);
      \draw [thick] (i2) -- (v1);

      \draw [thick,double] (v1) -- (v2);

      \draw [thick] (v2) -- (o1);
      \draw [thick] (v2) -- (o2);

      \fill (v1) circle [radius=0.1];
      \fill (v2) circle [radius=0.1];

      \node at (i1) [anchor=east] {\small $x_{\sL}^{\pm}$};
      \node at (i2) [anchor=west] {\small $y_{\sL}^{\pm}$};

      \node at (o1) [anchor=east] {\small $y_{\sL}^{\pm}$};
      \node at (o2) [anchor=west] {\small $x_{\sL}^{\pm}$};

      \node at ($0.5*(v1)+0.5*(v2)$) [anchor=west] {\small $z_{\sL}^{\pm}$};

      \node at ($0.5*(i1)+0.5*(v1)$) [anchor=south,rotate=45] {$\scriptscriptstyle M=+1$};
      \node at ($0.5*(i2)+0.5*(v1)$) [anchor=south,rotate=-45] {$\scriptscriptstyle M=+1$};
      \node at ($0.5*(v1)+0.5*(v2)$) [anchor=south,rotate=90] {$\scriptscriptstyle M=+2$};
      \node at ($0.5*(o1)+0.5*(v2)$) [anchor=south,rotate=-45] {$\scriptscriptstyle M=+1$};
      \node at ($0.5*(o2)+0.5*(v2)$) [anchor=south,rotate=45] {$\scriptscriptstyle M=+1$};

      \node at (2,0.75) [anchor=center] {\small $x_{\sL}^- = y_{\sL}^+$};
      \node at (2,0) [anchor=center] {\small $z_{\sL}^+ = x_{\sL}^+$};
      \node at (2,-0.75) [anchor=center] {\small $z_{\sL}^- = y_{\sL}^-$};
    \end{tikzpicture}
  }
  \subfloat[T channel\label{fig:AdS3-T-channel-LR}]{
    \begin{tikzpicture}
      \useasboundingbox (-3.5,-2.5) rectangle (3.5,2.5);

      \coordinate (i1) at (-1,-2);
      \coordinate (i2) at (+1,-2);

      \coordinate (o1) at (-1,+2);
      \coordinate (o2) at (+1,+2);

      \coordinate(v1) at (-1,-0.5);
      \coordinate (v2) at (+1,+0.5);

      \draw [thick] (i1) -- (v1);
      \draw [thick] (i2) -- (v2);

      \draw [thick,double] (v1) -- (v2);

      \draw [thick] (v1) -- (o1);
      \draw [thick] (v2) -- (o2);

      \fill (v1) circle [radius=0.1];
      \fill (v2) circle [radius=0.1];

      \node at (i1) [anchor=east] {\small $x_{\sL}^{\pm}$};
      \node at (i2) [anchor=west] {\small $y_{\sR}^{\pm}$};

      \node at (o1) [anchor=east] {\small $y_{\sR}^{\pm}$};
      \node at (o2) [anchor=west] {\small $x_{\sL}^{\pm}$};

      \node at ($0.5*(v1)+0.5*(v2)$) [anchor=north] {\small $z_{\sL}^{\pm}$};

      \node at (-1,-1.25) [anchor=south,rotate=90] {$\scriptscriptstyle M=+1$};
      \node at (+1,-1.25) [anchor=north,rotate=90] {$\scriptscriptstyle M=-1$};
      \node at (+1,+1.25) [anchor=north,rotate=90] {$\scriptscriptstyle M=+1$};
      \node at (-1,+1.25) [anchor=south,rotate=90] {$\scriptscriptstyle M=-1$};

      \node at (0,0) [anchor=south,rotate=26.5651] {$\scriptscriptstyle M=+2$};

      \node at (2.5,0.75) [anchor=center] {\small $\mathllap{x_{\sL}^-} = \mathrlap{1/y_{\sR}^+}$};
      \node at (2.5,0) [anchor=center] {\small $\mathllap{z_{\sL}^+} = \mathrlap{x_{\sL}^+}$};
      \node at (2.5,-0.75) [anchor=center] {\small $\mathllap{z_{\sL}^-} = \mathrlap{1/y_{\sR}^-}$};
    \end{tikzpicture}
  }
  \caption{S and T channel diagrams for $\AdS_3$. The diagrams are showing the LL and LR processes, respectively. The diagrams for RR and RL, take exactly the same form.}
\end{figure}
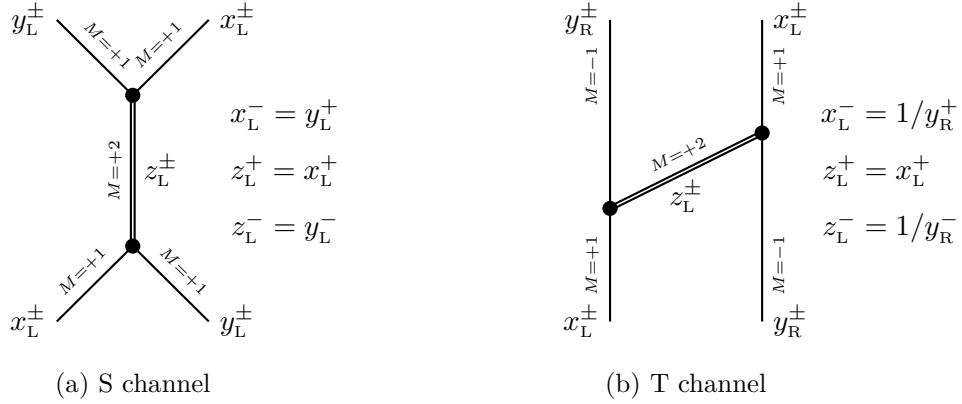

With the above remarks taken into account, the set of vertices we find in $\AdS_3\times\Sphere^3\times\Torus^4$ is closely related those of $\AdS_5\times\Sphere^5$. In summary, we find\footnote{
  In appendix~\ref{sec:matrix-representations} we give explicit matrix realisations of the corresponding bound state representations.
}
\begin{itemize}
\item An S channel pole in the massive LL and RR sectors, located at $x_1^+ = x_2^-$, as illustrated in figure~\ref{fig:AdS3-S-channel-LL}.
\item A T channel pole in the massive LR and RL sectors, located at $x_1^+ = 1/x_2^-$, as illustrated in figure~\ref{fig:AdS3-T-channel-LR}.
\item Double poles in all sectors involving only massive excitations, \ie, LL, LR, RL and RR, located at
  \begin{equation}
    u_1 - u_2 = - \frac{2in}{h}, \qquad n \in \Integers, \qquad n > 1 .
    \label{eq:dhm-dbl-ads3}
  \end{equation}
\end{itemize}
Let us see in more detail how the double poles appear. The relevant diagrams for LL scattering are shown in figures~\ref{fig:AdS3-box-LL-LLLR} and~\ref{fig:AdS3-box-LL-RRRL}. The external legs are the same in the two diagrams, while the bound states in the loop run in the opposite directions. For each diagram there are two ways to assign the four vertices.\footnote{
  For each vertex we found two solution to the energy, momentum and charge conservation, so \textit{\`a priori} there are 16 possible combinations for the box diagrams in question. However, all but two lead to lead to over constrained sets of equations.
}
For diagram~\ref{fig:AdS3-box-LL-RRRL} we find
\begin{equation}
  x_{1\sL}^+ = \frac{1}{z_{1\sR}^+} = y_{2\sL}^+ , \quad
  x_{1\sL}^- = \frac{1}{z_{2\sR}^+} = \frac{1}{y_{1\sR}^+} , \quad
  x_{2\sL}^+ = \frac{1}{z_{1\sR}^-} = \frac{1}{y_{1\sR}^-}, \quad
  x_{2\sL}^- = \frac{1}{z_{2\sR}^-} = y_{2\sR}^-, 
\end{equation}
or
\begin{equation}
  x_{1\sL}^+ = \frac{1}{z_{2\sR}^-} = \frac{1}{y_{1\sR}^-} , \quad
  x_{1\sL}^- = \frac{1}{z_{1\sR}^-} = y_{2\sL}^- , \quad
  x_{2\sL}^+ = \frac{1}{z_{2\sR}^+} = y_{2\sL}^+, \quad
  x_{2\sL}^- = \frac{1}{z_{1\sR}^+} = \frac{1}{y_{1\sR}^+} ,
\end{equation}
which leads to
\begin{equation}
  u_1 - u_2 = +\frac{2in}{h} , \qquad \text{or} \qquad u_1 - u_2 = -\frac{2in}{h} .
\end{equation}
Since we assume that there is no vertex involving two massive and one massless excitation, we get the condition $n > 1$. As before, we obtain a physical pole by demanding that the imaginary part of the  momentum of the excitation $x_{1\sL}^{\pm}$ is positive, which selects the second solution as indicated in equation~\eqref{eq:dhm-dbl-ads3}.\footnote{
  To see this we note that $\partial_p u(p) < 0$ for real momentum $p$. Writing $u(p+iq) - u(p-iq) \approx 2iq\partial_p u(p)$ we find that the solution with a negative imaginary part of $u_1 - u_2$ corresponds to the physical pole.
}

Similarly, demanding that diagram~\ref{fig:AdS3-box-LL-LLLR} leads to a physical pole leads to
\begin{equation}
  x_{1\sL}^+ = z_{1\sL}^- = \frac{1}{y_{2\sR}^-} , \quad
  x_{1\sL}^- = z_{2\sL}^- = y_{1\sL}^- , \quad
  x_{2\sL}^+ = z_{1\sL}^+ = y_{1\sL}^+ , \quad
  x_{2\sL}^- = z_{2\sL}^+ = \frac{1}{y_{2\sR}^+} ,
\end{equation}
from which we again obtain the condition~\eqref{eq:dhm-dbl-ads3}.

Diagrams~\ref{fig:AdS3-box-LR-LLLR} and~\ref{fig:AdS3-box-LR-RRRL} give rise to double poles in the LR S matrix. Physical poles are obtained when
\begin{equation}
  x_{1\sL}^+ = z_{1\sL}^- = \frac{1}{y_{2\sR}^-} , \quad
  x_{1\sL}^- = z_{2\sL}^- = y_{1\sL}^- , \quad
  \frac{1}{x_{2\sR}^+} = z_{2\sL}^+ = \frac{1}{y_{2\sR}^+} , \quad
  \frac{1}{x_{2\sR}^-} = z_{1\sL}^+ = y_{1\sL}^+ ,
\end{equation}
and
\begin{equation}
  x_{1\sL}^+ = \frac{1}{z_{2\sR}^-} = \frac{1}{y_{1\sR}^-} , \quad
  x_{1\sL}^- = \frac{1}{z_{1\sR}^-} = y_{2\sL}^- , \quad
  \frac{1}{x_{2\sR}^+} = \frac{1}{z_{1\sR}^+} = \frac{1}{y_{1\sR}^+} , \quad
  \frac{1}{x_{2\sR}^-} = z_{2\sL}^+ = \frac{1}{y_{2\sL}^+} ,
\end{equation}
respectively. Again these two sets of equations give rise to the condition~\eqref{eq:dhm-dbl-ads3}.

In the next section we will show that this set of simple and double poles is consistent with the dressing phases proposed in~\cite{Borsato:2013hoa,Borsato:2016kbm,Borsato:2016xns}.

\begin{figure}
  \centering

  \subfloat[\label{fig:AdS3-box-LL-RRRL}]{
    \begin{tikzpicture}
      \coordinate (i1) at (-2.75,-2.5);
      \coordinate (i2) at (+2,-2.5);

      \coordinate (o1) at (-2.75,+2.5);
      \coordinate (o2) at (+2,+2.5);

      \coordinate (v1) at (-1,-0.75);
      \coordinate (v2) at (+1,-1.5);
      \coordinate (v3) at (+1,+1.5);
      \coordinate (v4) at (-1,+0.75);

      \draw [thick] (i1) -- (v1);
      \draw [thick] (i2) -- (v2);
      \draw [thick] (v3) -- (o2);
      \draw [thick] (v4) -- (o1);

      \draw [thick, double] (v1) -- (v2) -- (v3) -- (v4) -- (v1);

      \fill (v1) circle [radius=0.1];
      \fill (v2) circle [radius=0.1];
      \fill (v3) circle [radius=0.1];
      \fill (v4) circle [radius=0.1];

      \node at (i1) [anchor=east] {\small $x_{1\sL}^{\pm}$};
      \node at (i2) [anchor=west] {\small $x_{2\sL}^{\pm}$};
      \node at (o1) [anchor=east] {\small $x_{2\sL}^{\pm}$};
      \node at (o2) [anchor=west] {\small $x_{1\sL}^{\pm}$};

      \node at ($0.5*(v1)+0.5*(v2)$) [anchor=north] {\small $z_{1\sR}^{\pm}$};
      \node at ($0.5*(v2)+0.5*(v3)$) [anchor=west] {\small $y_{2\sL}^{\pm}$};
      \node at ($0.5*(v3)+0.5*(v4)$) [anchor=south] {\small $z_{2\sR}^{\pm}$};
      \node at ($0.5*(v4)+0.5*(v1)$) [anchor=east] {\small $y_{1\sR}^{\pm}$};

      \node at ($(i1)+(+0.5,+0.5)$) [anchor=south,rotate=+45] {$\scriptscriptstyle M=+1$};
      \node at ($(i2)+(-0.5,+0.5)$) [anchor=south,rotate=-45] {$\scriptscriptstyle M=+1$};
      \node at ($(o1)-(-0.5,+0.5)$) [anchor=south,rotate=-45] {$\scriptscriptstyle M=+1$};
      \node at ($(o2)-(+0.5,+0.5)$) [anchor=south,rotate=+45] {$\scriptscriptstyle M=+1$};

      \node at ($0.5*(v1)+0.5*(v4)$) [anchor=north,rotate=90] {$\scriptscriptstyle M=-n+1$};
      \node at ($0.5*(v2)+0.5*(v3)$) [anchor=south,rotate=90] {$\scriptscriptstyle M=n+1$};
      \node at ($0.5*(v1)+0.5*(v2)$) [anchor=south,rotate=-20.556] {$\scriptscriptstyle M=-n$};
      \node at ($0.5*(v3)+0.5*(v4)$) [anchor=north,rotate=+20.556] {$\scriptscriptstyle M=-n$};
    \end{tikzpicture}
  }
  \hspace*{1cm}
  \subfloat[\label{fig:AdS3-box-LL-LLLR}]{
    \begin{tikzpicture}
      \coordinate (i1) at (-2.75,-2.5);
      \coordinate (i2) at (+2,-2.5);

      \coordinate (o1) at (-2.75,+2.5);
      \coordinate (o2) at (+2,+2.5);

      \coordinate (v1) at (-1,-0.75);
      \coordinate (v2) at (+1,-1.5);
      \coordinate (v3) at (+1,+1.5);
      \coordinate (v4) at (-1,+0.75);

      \draw [thick] (i1) -- (v1);
      \draw [thick] (i2) -- (v2);
      \draw [thick] (v3) -- (o2);
      \draw [thick] (v4) -- (o1);

      \draw [thick, double] (v1) -- (v2) -- (v3) -- (v4) -- (v1);

      \fill (v1) circle [radius=0.1];
      \fill (v2) circle [radius=0.1];
      \fill (v3) circle [radius=0.1];
      \fill (v4) circle [radius=0.1];

      \node at (i1) [anchor=east] {\small $x_{1\sL}^{\pm}$};
      \node at (i2) [anchor=west] {\small $x_{2\sL}^{\pm}$};
      \node at (o1) [anchor=east] {\small $x_{2\sL}^{\pm}$};
      \node at (o2) [anchor=west] {\small $x_{1\sL}^{\pm}$};

      \node at ($0.5*(v1)+0.5*(v2)$) [anchor=north] {\small $z_{1\sL}^{\pm}$};
      \node at ($0.5*(v2)+0.5*(v3)$) [anchor=west] {\small $y_{2\sR}^{\pm}$};
      \node at ($0.5*(v3)+0.5*(v4)$) [anchor=south] {\small $z_{2\sL}^{\pm}$};
      \node at ($0.5*(v4)+0.5*(v1)$) [anchor=east] {\small $y_{1\sL}^{\pm}$};

      \node at ($(i1)+(+0.5,+0.5)$) [anchor=south,rotate=+45] {$\scriptscriptstyle M=+1$};
      \node at ($(i2)+(-0.5,+0.5)$) [anchor=south,rotate=-45] {$\scriptscriptstyle M=+1$};
      \node at ($(o1)-(-0.5,+0.5)$) [anchor=south,rotate=-45] {$\scriptscriptstyle M=+1$};
      \node at ($(o2)-(+0.5,+0.5)$) [anchor=south,rotate=+45] {$\scriptscriptstyle M=+1$};

      \node at ($0.5*(v1)+0.5*(v4)$) [anchor=north,rotate=90] {$\scriptscriptstyle M=n+1$};
      \node at ($0.5*(v2)+0.5*(v3)$) [anchor=south,rotate=90] {$\scriptscriptstyle M=-n+1$};
      \node at ($0.5*(v1)+0.5*(v2)$) [anchor=south,rotate=-20.556] {$\scriptscriptstyle M=n$};
      \node at ($0.5*(v3)+0.5*(v4)$) [anchor=north,rotate=+20.556] {$\scriptscriptstyle M=n$};
    \end{tikzpicture}
  }
\\[1cm]
  \subfloat[\label{fig:AdS3-box-LR-LLLR}]{
    \begin{tikzpicture}
      \coordinate (i1) at (-2.75,-2.5);
      \coordinate (i2) at (+2,-2.5);

      \coordinate (o1) at (-2.75,+2.5);
      \coordinate (o2) at (+2,+2.5);

      \coordinate (v1) at (-1,-0.75);
      \coordinate (v2) at (+1,-1.5);
      \coordinate (v3) at (+1,+1.5);
      \coordinate (v4) at (-1,+0.75);

      \draw [thick] (i1) -- (v1);
      \draw [thick] (i2) -- (v2);
      \draw [thick] (v3) -- (o2);
      \draw [thick] (v4) -- (o1);

      \draw [thick, double] (v1) -- (v2) -- (v3) -- (v4) -- (v1);

      \fill (v1) circle [radius=0.1];
      \fill (v2) circle [radius=0.1];
      \fill (v3) circle [radius=0.1];
      \fill (v4) circle [radius=0.1];

      \node at (i1) [anchor=east] {\small $x_{1\sL}^{\pm}$};
      \node at (i2) [anchor=west] {\small $x_{2\sR}^{\pm}$};
      \node at (o1) [anchor=east] {\small $x_{2\sR}^{\pm}$};
      \node at (o2) [anchor=west] {\small $x_{1\sL}^{\pm}$};

      \node at ($0.5*(v1)+0.5*(v2)$) [anchor=north] {\small $z_{1\sL}^{\pm}$};
      \node at ($0.5*(v2)+0.5*(v3)$) [anchor=west] {\small $y_{2\sR}^{\pm}$};
      \node at ($0.5*(v3)+0.5*(v4)$) [anchor=south] {\small $z_{2\sL}^{\pm}$};
      \node at ($0.5*(v4)+0.5*(v1)$) [anchor=east] {\small $y_{1\sL}^{\pm}$};

      \node at ($(i1)+(+0.5,+0.5)$) [anchor=south,rotate=+45] {$\scriptscriptstyle M=+1$};
      \node at ($(i2)+(-0.5,+0.5)$) [anchor=south,rotate=-45] {$\scriptscriptstyle M=-1$};
      \node at ($(o1)-(-0.5,+0.5)$) [anchor=south,rotate=-45] {$\scriptscriptstyle M=-1$};
      \node at ($(o2)-(+0.5,+0.5)$) [anchor=south,rotate=+45] {$\scriptscriptstyle M=+1$};

      \node at ($0.5*(v1)+0.5*(v4)$) [anchor=north,rotate=90] {$\scriptscriptstyle M=+n$};
      \node at ($0.5*(v2)+0.5*(v3)$) [anchor=south,rotate=90] {$\scriptscriptstyle M=-n$};
      \node at ($0.5*(v1)+0.5*(v2)$) [anchor=south,rotate=-20.556] {$\scriptscriptstyle M=n-1$};
      \node at ($0.5*(v3)+0.5*(v4)$) [anchor=north,rotate=+20.556] {$\scriptscriptstyle M=n+1$};
    \end{tikzpicture}
  }
  \hspace*{1cm}
  \subfloat[\label{fig:AdS3-box-LR-RRRL}]{
    \begin{tikzpicture}
      \coordinate (i1) at (-2.75,-2.5);
      \coordinate (i2) at (+2,-2.5);

      \coordinate (o1) at (-2.75,+2.5);
      \coordinate (o2) at (+2,+2.5);

      \coordinate (v1) at (-1,-0.75);
      \coordinate (v2) at (+1,-1.5);
      \coordinate (v3) at (+1,+1.5);
      \coordinate (v4) at (-1,+0.75);

      \draw [thick] (i1) -- (v1);
      \draw [thick] (i2) -- (v2);
      \draw [thick] (v3) -- (o2);
      \draw [thick] (v4) -- (o1);

      \draw [thick, double] (v1) -- (v2) -- (v3) -- (v4) -- (v1);

      \fill (v1) circle [radius=0.1];
      \fill (v2) circle [radius=0.1];
      \fill (v3) circle [radius=0.1];
      \fill (v4) circle [radius=0.1];

      \node at (i1) [anchor=east] {\small $x_{1\sL}^{\pm}$};
      \node at (i2) [anchor=west] {\small $x_{2\sR}^{\pm}$};
      \node at (o1) [anchor=east] {\small $x_{2\sR}^{\pm}$};
      \node at (o2) [anchor=west] {\small $x_{1\sL}^{\pm}$};

      \node at ($0.5*(v1)+0.5*(v2)$) [anchor=north] {\small $z_{1\sR}^{\pm}$};
      \node at ($0.5*(v2)+0.5*(v3)$) [anchor=west] {\small $y_{2\sL}^{\pm}$};
      \node at ($0.5*(v3)+0.5*(v4)$) [anchor=south] {\small $z_{2\sR}^{\pm}$};
      \node at ($0.5*(v4)+0.5*(v1)$) [anchor=east] {\small $y_{1\sR}^{\pm}$};

      \node at ($(i1)+(+0.5,+0.5)$) [anchor=south,rotate=+45] {$\scriptscriptstyle M=+1$};
      \node at ($(i2)+(-0.5,+0.5)$) [anchor=south,rotate=-45] {$\scriptscriptstyle M=-1$};
      \node at ($(o1)-(-0.5,+0.5)$) [anchor=south,rotate=-45] {$\scriptscriptstyle M=-1$};
      \node at ($(o2)-(+0.5,+0.5)$) [anchor=south,rotate=+45] {$\scriptscriptstyle M=+1$};

      \node at ($0.5*(v1)+0.5*(v4)$) [anchor=north,rotate=90] {$\scriptscriptstyle M=-n$};
      \node at ($0.5*(v2)+0.5*(v3)$) [anchor=south,rotate=90] {$\scriptscriptstyle M=+n$};
      \node at ($0.5*(v1)+0.5*(v2)$) [anchor=south,rotate=-20.556] {$\scriptscriptstyle M=-n-1$};
      \node at ($0.5*(v3)+0.5*(v4)$) [anchor=north,rotate=+20.556] {$\scriptscriptstyle M=-n+1$};
    \end{tikzpicture}
  }
  
  \caption{Box diagrams for LL and LR scattering leading to double poles in the S matrix. We indicate explicitly the masses $M$ of the excitations, with each diagram corresponding to a family of processes labeled by an integer $n>1$.}
\end{figure}

\section{Comparison with the $\AdS_3$ dressing phase}
\label{sec:AdS3-comparison}

The symmetries of the theory determined the world-sheet S matrix of $\AdS_3\times \Sphere^3\times\Torus^4$ up to four dressing factors: $\sigma^{\bullet\bullet},{\tilde \sigma}^{\bullet\bullet},\sigma^{\bullet\circ},\sigma^{\circ\circ}$ all of which satisfy known crossing equations~\cite{Borsato:2013qpa,Borsato:2014exa,Borsato:2014hja}. The former two dressing factors enter the scattering of massive LL and LR excitations, respectively, while the latter two involve mixed-mass and massless scattering. It is convenient to express the dressing factors as phases, for example
\begin{equation}
\sigma^{\bullet\bullet}(p_1,p_2)=e^{i\theta^{\bullet\bullet}(p_1,p_2)},
\end{equation}
with similar expressions for the other $\sigma$s. From the structure of higher-conserved charges in the system, it is known that phases $\theta^{\bullet\bullet}$ can be decomposed as~\cite{Arutyunov:2004vx}
\begin{equation}
\theta^{\bullet\bullet}=\chi(x^+_1,x^+_2)-\chi(x^-_1,x^+_2)-\chi(x^+_1,x^-_2)+\chi(x^-_1,x^-_2).
\label{eq:chi-decomp}
\end{equation}
Analogous decompositions hold for the other $\theta$s. The $\chi$ satisfy crossing equations that follow from the $\sigma$ ones but are simpler. Solutions for $\chi^{\bullet\bullet}$ and 
${\tilde\chi}^{\bullet\bullet}$ were found in~\cite{Borsato:2013hoa} and in the physical $|x_i^\pm|>1$ region they take the form
\begin{align}
  \chi^{\bullet\bullet}(x_1,x_2) &= \chi^{\text{BES}}(x_1,x_2)+\tfrac{1}{2}\left(-\chi^{\text{HL}}(x_1,x_2)+\chi^{-}(x_1,x_2)\right) , \\
  \widetilde{\chi}^{\bullet\bullet}(x_1,x_2) &= \chi^{\text{BES}}(x_1,x_2)+\tfrac{1}{2}\left(-\chi^{\text{HL}}(x_1,x_2)-\chi^{-}(x_1,x_2)\right) ,
\label{eq:ads3-bb-chi-solution}
\end{align}
where  $\chi^{\text{BES}}$ is the Beisert-Eden-Staudacher (BES) phase~\cite{Beisert:2006ez}, which can be expressed as a double-contour integral~\cite{Dorey:2007xn}
\begin{equation}
\chi^\BES(x_1,x_2)= i \oint \frac{dw_1}{2 \pi i} \oint \frac{dw_2}{2 \pi i} \, \frac{1}{x_1-w_1}\frac{1}{x_2-w_2} \log{\frac{\Gamma[1+\frac{ih}{2}(w_1+1/w_1-w_2-1/w_2)]}{\Gamma[1-\frac{ih}{2}(w_1+1/w_1-w_2-1/w_2)]}},
\label{eq:besdhmrep}
\end{equation}
and
\begin{align}
\label{eq:intu-intd-HL}
\chi^\HL(x_1,x_2)&=\left( \inturl - \intdlr \right)\frac{dw}{4\pi} \frac{1}{x_1-w} \left(\log{(x_2-w)}-\log{\left(x_2-1/w\right)}\right),
\\
\label{eq:chi-}
\chi^-(x_1,x_2) &=\left( \inturl - \intdlr \right)\frac{dw}{8\pi} \frac{1}{x_1-w} 
\log{\left[ (x_2-w)\left(1-\frac{1}{x_2w}\right)\right]} \ - \left(x_1 \longleftrightarrow x_2\right).
\end{align}
We see that $\chi^{\bullet\bullet}$ and ${\tilde\chi}^{\bullet\bullet}$ are different from the $\AdS_5\times\Sphere^5$ BES phase, but that the modification is relatively simple and involves only terms at the Hern\'andez-L\'opez order~\cite{Hernandez:2006tk}. Compared to the BES phase, these extra terms have a simple analytic structure, which we will analyse later on in this section. 

\subsection{DHM double poles from the BES dressing factor}

First, let us briefly recall how DHM double poles appear in the BES phase.\footnote{See~\cite{Vieira:2010kb} for a review of the dressing factor and its analytic structure based on the elegant derivation~\cite{Volin:2009uv}.} At the level of the $\chi$, crossing from the physical $|x_i|>1$ region amounts to analytically continuing $x_i$ inside the unit disc. In doing that, $x_i$ moves from one side of the contours that define the $\chi$s above to the other. The Sochocki-Plemelj theorem then dictates that the analytically continued $\chi$ pick up a term proportional to the residue of the integrand. For the BES phase inside the singly-crossed region we find
\begin{align}
\chi_{\text{cr}}^\BES(x_1,x_2)=& \chi^\BES(x_1,x_2)
+ i  \oint \frac{dw_2}{2 \pi i} \, \frac{1}{x_2-w_2} \log{\frac{\Gamma\left[1+\frac{ih}{2}(x_1+\tfrac{1}{x_1}-w_2-\tfrac{1}{w_2})\right]}{\Gamma\left[1-\frac{ih}{2}(x_1+\tfrac{1}{x_1}-w_2-\tfrac{1}{w_2})\right]}} ,
\label{eq:bes-crossed}
\end{align}
where $\chi^\BES_{\text{cr}}$  is the analytic continuation of the integral in equation~\eqref{eq:besdhmrep} into the region $|x_1|<1$. The first term on the right hand side above is given by the integral in equation~\eqref{eq:besdhmrep} now evaluated for $|x_1|<1$. The second term is needed to ensure continuity across the unit circle $|x_1|=1$ since the integral~\eqref{eq:besdhmrep} is discontinuous there. This term has important physical consequences: it introduces new cuts inside the unit disc, see for example Figure 2 of~\cite{Arutyunov:2009kf}. To see this explicitly, we integrate it by parts to get
\begin{equation}
  \begin{aligned}
    h \oint \frac{dw_2}{2 \pi i} \, \left(1-\tfrac{1}{w_2^2}\right)\log(x_2-w_2) 
    &\Bigl[\psi\bigl(1+\tfrac{i h}{2}(x_1+\tfrac{1}{x_1}-w_2-\tfrac{1}{w_2})\bigr)
    \\
    & \quad
      +\psi\bigl(1-\tfrac{i h}{2}(x_1+\tfrac{1}{x_1}-w_2-\tfrac{1}{w_2})\bigr)\Bigr],
  \end{aligned}
  \label{eq:psi-fn-pbp}
\end{equation}
where $\psi(x)\equiv \Gamma'(x)/\Gamma(x)$ is the digamma function. Since $\psi(x)$ has poles for $x$ a negative integer, the last term on the right hand side of equation~\eqref{eq:bes-crossed} has cuts when
\begin{equation}
x_1+\tfrac{1}{x_1}-w_2-\tfrac{1}{w_2} =\tfrac{2i}{h}n,\qquad n\in\Integers \,,\qquad n\neq 0
\label{eq:dhm-branch-cuts-1}
\end{equation}
The above equations are invariant under $w_2\rightarrow \tfrac{1}{w_2}$, hence each one gives rise to two poles at antipodal points of the unit circle spanned by $w_2$. Analytically continuing through one of these cuts requires us to further modify the expression for $\chi_{\text{cr}}^\BES$ . In particular, as a consequence of the Sochocki-Plemelj theorem, the analytic continuation through the $n$-th cut of the term in equation~\eqref{eq:psi-fn-pbp} modifies the integral expression for $\chi_{\text{cr}}^\BES$ by an extra term
\begin{equation}
-i\,\mbox{sign}(n)\log\bigl(u(x_2)-u(x_1)+i n\bigr) ,
\label{eq:dhm-dbl-pole}
\end{equation}
as reviewed in equation (3.23) of~\cite{Vieira:2010kb}. This term leads to double poles and zeros in the S matrix dressing factor $\sigma^2=e^{2i\theta}$ whose location~\cite{Dorey:2007xn} agrees precisely with those found in section~\ref{sssec:double-poles}.\footnote{In the analysis leading up to equation~\eqref{eq:dhm-dbl-pole}, we were already in the $x_1$ crossed region, and as such we should not cross the $n=1$ cut in~\eqref{eq:dhm-branch-cuts-1}.}

\subsection{
\texorpdfstring{Analytic structure of $\chi^{\bullet\bullet}$ and $\tilde{\chi}^{\bullet\bullet}$}{Analytic structure of $\chi^{\bullet\bullet}$ and $\chi^{\bullet\bullet}$}}
\label{sec:massive-anal-str}

The dressing phases $\chi^{\bullet\bullet}$ and ${\tilde\chi}^{\bullet\bullet}$ are defined in the physical region $|x_i|>1$ in equation~\eqref{eq:ads3-bb-chi-solution}. Their continuation inside the unit circle was analysed in~\cite{Borsato:2013hoa}. There it was shown that the HL-order terms after crossing in $x_1$ take the form\footnote{See equations (A.7) and (A.25) of~\cite{Borsato:2013hoa}.}
\begin{align}
  \label{eq:chi-mi-cr}
  \chi_{\text{cr}}^-(x_1,x_2)
  &= \chi^-(x_1,x_2)-\frac{i}{2}\,\log\Bigl[(x_2-x_1)\bigl(1-\tfrac{1}{x_1x_2}\bigr)\Bigr] ,
  \\
  \label{eq:chi-hl-cr}
  \chi_{\text{cr}}^\HL(x_1,x_2)
  &= \chi^\HL(x_1,x_2)-\frac{i}{2}\,\log\frac{x_2-x_1}{x_2-1/x_1} .
\end{align}
The first terms on the right hand side of the above equations are given by the integrals in~\eqref{eq:intu-intd-HL}
and~\eqref{eq:chi-}, respectively, now evaluated with $|x_1|<1$. The final terms in the above two equations modify the dressing factors $e^{2i\chi^{\bullet\bullet}}$ and $e^{2i{\tilde \chi}^{\bullet\bullet}}$  that enter the S matrix by a \textit{rational} function of $x_i$.  Dropping terms dependent on $x_2$ only, which do not contribute to the final expressions for $\theta$, the dressing factors in the $x_1$-crossed region are
\begin{equation}
e^{2i\chi_{\text{cr}}^{\bullet\bullet}}=\left(x_2-\frac{1}{x_1}\right)e^{2i\chi^{\bullet\bullet}},\qquad
e^{2i{\tilde\chi}_{\text{cr}}^{\bullet\bullet}}=\frac{1}{x_2-x_1}
e^{2i{\tilde\chi}^{\bullet\bullet}}.
\label{eq:mive-crossed}
\end{equation}
We see that the analytic structure inside the unit circle $|x_1|<1$ is determined by $\chi^\BES$ as discussed in the previous sub-section. The HL-order terms $\chi^-_{\text{cr}}$ and $\chi^\HL_{\text{cr}}$ given in equations~\eqref{eq:chi-mi-cr} and~\eqref{eq:chi-hl-cr}, do not introduce any new cuts and lead to \textit{analytic} modifications of the dressing factor $\sigma^2_\BES$. This shows that 
$(\sigma^{\bullet\bullet})^2$ and $({\tilde \sigma}^{\bullet\bullet})^2$, the massive dressing phases of the R-R $\AdS^3\times\Sphere^3\times\Torus^4$ S matrix, have  double poles and zeros whose location~\eqref{eq:dhm-dbl-pole} is the same as that of the DHM double poles and zeros in $\AdS_5\times\Sphere^5$.

\subsection{
\texorpdfstring{Analytic structure of $\chi^{\circ\bullet}$ and $\chi^{\circ\circ}$}{Analytic structure of $\chi^{\bullet\circ}$ and $\chi^{\circ\circ}$}}

In~\cite{Borsato:2016kbm,Borsato:2016xns,Fontanella:2019ury}, expressions for  $\chi^{\circ\bullet}$ and $\chi^{\circ\circ}$ were proposed
\begin{align}
  \theta^{\circ\bullet}(x_1,x_2)
  &= \left[\theta^\AFS(x_1,x_2)+\tfrac{1}{2}\theta^\HL(x_1,x_2)\right]_{m_x=0,\,m_y=1},
  \\
  \theta^{\circ\circ}(x_1,x_2)
  &=\left[\tfrac{1}{2}\theta^\HL(x_1,x_2)\right]_{m_x=m_y=0}.
\end{align}
In the physical region, the HL-order terms are given by deforming the contour of integration in equation~\eqref{eq:intu-intd-HL} to the interval $[-1,\,1]$ as described in detail in section 2.3.3 of~\cite{Borsato:2016xns}, while $\theta^\AFS$ is constructed from
\begin{equation}
  \chi^\AFS(x_1,x_2)=\tfrac{1}{x_1}-\tfrac{1}{x_2}+\left(x_2+\tfrac{1}{x_2}-x_1-\tfrac{1}{x_1}\right)\log\left(1-\tfrac{1}{x_1x_2}\right).
\end{equation}
In massless kinematics the AFS-order term is trivial under crossing and so is a potential CDD factor,
whose presence was not excluded in~\cite{Borsato:2016kbm,Borsato:2016xns}. In~\cite{Fontanella:2019ury}, however, it was shown that such a term cannot be written in difference form in terms of the rapidity variable $\gamma\equiv \log\tan(p/4)$ introduced in~\cite{Fontanella:2019baq}. As a result, we do not include the massless AFS term. 

Analytically continuing the HL-order term to the crossed region modifies the dressing factors $(\sigma^{\circ\bullet})^2$ and $(\sigma^{\circ\circ})^2$  by \textit{rational} terms, entirely analogously to equation~\eqref{eq:mive-crossed}. Explicitly, from equations (2.47) and (2.48) in~\cite{Borsato:2016xns}, we find that the mixed-mass dressing factor in the massless crossed region is given by
\begin{equation}
  e^{2i\theta^{\circ\bullet}_{\text{cr}}(x_1\,,\,x_2^\pm)}=\frac{(x_1 x_2^+-1)(x_1-x_2^-)}{(x_1-x_2^+)(x_1 x_2^--1)} e^{2i\theta^{\circ\bullet}(x_1\,,\,x_2^\pm)},
\end{equation}
where $x_1\equiv x^+_1$ for the massless variable. Similarly, in the massive crossed region we find
\begin{equation}
  e^{2i\theta^{\bullet\circ}_{\text{cr}}(x^\pm_1\,,\,x_2)}=\frac{(x^+_1 x_2-1)(x^-_1-x_2)}{x^2_2(x^+_1-x_2)(x^-_1 x_2-1)} e^{2i\theta^{\bullet\circ}(x^\pm_1\,,\,x_2)},
\end{equation}
where now $x_2\equiv x_2^+$ is the massless variable.\footnote{To similify the notation we have always crossed in the \textit{first} argument of the dressing factor and so considered $\theta^{\circ\bullet}$ and $\theta^{\bullet\circ}$, respectively for the two types of mixed-mass crossed regions.} 

From the above equations we see that the mixed-mass dressing factors have a much simpler analytic structure than the massive dressing factors. As we discussed in section~\ref{sec:massive-anal-str}, the analytic structure of $\chi^{\bullet\bullet}$ and ${\tilde\chi}^{\bullet\bullet}$ in the crossed region is substantially modified by the second term on the right hand side of equation~\eqref{eq:bes-crossed}.
It is this more complicated analytic structure that leads to the DHM double poles in the massive dressing phases. The mixed mass dressing factor on the other hand is modified only by rational terms in the crossed region. As a result, it gives no new cuts or double poles, in agreement with the dynamical arguments presented in section~\ref{sec:expected-poles}.

The massless dressing factor in the $x_1$-crossed region takes the form
\begin{equation}
  e^{2i\theta^{\circ\circ}_{\text{cr}}(x_1\,,\,x_2)}=\left(\frac{x_1 x_2-1}{x_1-x_2}\right)^2 e^{2i\theta^{\circ\circ}(x_1\,,\,x_2)}.
\end{equation}
We again see that no new cuts are introduced.

\section{Conclusions}
\label{sec:conclusions}

In this paper we have investigated the analytic structure of the exact worldsheet S matrix of strings on $\AdS_3 \times \Sphere^3 \times \Torus^4$ with R-R flux. The bound states of the theory in the semi-classical regime are given by dyonic giant magnon solutions on $\Sphere^3$ much like in $\AdS_5\times\Sphere^5$~\cite{Dorey:2006dq}. However, the reduced dimensionality of the sphere in the $\AdS_3$ background leads to \textit{two} types of dyonic magnons, which we labelled L and R. These arise as bound states of $M=1$ and $M=-1$ fundamental excitations, respectively and transform in short representations of the off-shell symmetry algebra of the gauge-fixed theory~\cite{Borsato:2013hoa}. As in $\AdS_5$, it is expected that no other bound states exist.

The presence of bound states in the physical spectrum of a theory leads to poles and zeros in the S matrix. We showed that in the R-R $\AdS_3 \times \Sphere^3 \times \Torus^4$ background the bound states require simple and double poles, generalising the $\AdS_5\times\Sphere^5$ analysis of~\cite{Dorey:2007xn}. Using Landau diagrams, we determined the location of the simple poles in equation~\eqref{eq:T-pole-psu22-su2} and~\eqref{eq:S-pole-psu22-su2} and the double poles in equation~\eqref{eq:dhm-dbl-ads3}. We then showed that the exact S matrix proposed for this theory in~\cite{Borsato:2014exa,Borsato:2014hja} together with the dressing factors found in~\cite{Borsato:2013hoa,Borsato:2016kbm,Borsato:2016xns} has precisely the expected simple and double poles. The  location of the simple poles had already been checked in~\cite{Borsato:2013hoa}, since this essentially constitutes a consistency check on the solutions of the crossing equations. 

The location of DHM double poles places a much more stringent restriction on the S matrix. As we have shown in Appendix~\ref{sec:naturalness} one can find minimal solutions of the crossing equations (see equation~\eqref{eq:soln-hl-ord-exact}) by judicious normalization choices of the rational part of the S matrix~\eqref{eq:min-norm-choice}. The resulting dressing factors are independent of the coupling constant $h$, in other words come in at the Hern\'andez-L\'opez order~\cite{Hernandez:2006tk} and have the correct simple poles and zeros. However, due to their relatively simple analytic structure (essentially these are dilogarithms),  they do not have any DHM double poles. One then needs to add CDD-type homogeneous solutions of the crossing equations to obtain these double poles as in equation~\eqref{eq:soln-hom}. Given the intricate analytic structure of these homogeneous solutions, it would have been very challenging to find them \textit{a priori}, providing further motivation for normalising the rational part of the S matrix as in~\cite{Borsato:2014hja}.

In addition to the $M=\pm 1$ excitations, the $\AdS_3$ worldsheet theory contains massless modes~\cite{Sax:2012jv}, which do not form bound states~\cite{Borsato:2013hoa}. As discussed, in section~\ref{sec:DHM-AdS3}, the kinematics of the short $\alg{psu}(1|1)^4_{\mbox{c.e.}}$ multiplets \textit{does} allow for massless bound states to form from L and an R fundamental excitations in the T channel. In the physical theory these would have to be interpreted as fundamental $M=0$ excitations. As we showed, such kinematically allowed massless bound states need to be excluded since they have the wrong statistics and do not carry $\alg{su}(2)_\circ$ charges expected of fundamental massless excitations.

The $\AdS_3\times\Sphere^3\times\Torus^4$ S matrix has dressing factors associated to mixed mass and massless scattering processes. Solutions of the corresponding crossing equations were proposed in~\cite{Borsato:2016kbm,Borsato:2016xns} and an elegant form of the massless dressing factor, based on the $\gamma$ rapidity~\cite{Fontanella:2019baq}, was found in~\cite{Bombardelli:2018jkj,Fontanella:2019ury}.\footnote{Remarkably, the massless dressing factor~\cite{Borsato:2016kbm,Borsato:2016xns} turns out to be exactly the same as the famous soliton-anti-soliton scattering in sine Gordon theory found by Zamolodchikovs~\cite{Zamolodchikov:1978xm} at $\beta^2=16\pi/3$.} We showed that the analytic structure of massless and mixed mass dressing factors is much simpler than of the massive ones. They closely resemble the solutions discussed in Appendix~\ref{sec:naturalness} with dilogarithm-like cuts and no simple poles or zeros in the physical strip, nor any DHM poles, in agreement with the Landau-diagram expectations.

In summary, the analytic properties of the R-R $\AdS_3\times\Sphere^3\times\Torus^4$ S matrix combine features of two very different types of integrable theories. On the one hand, the massive S matrix exhibits an intricate analytic structure typical of non-local spin-chains that appear in higher-dimensional holographic models~\cite{Beisert:2006ez}. On the other hand, the mixed-mass and massless S matrices resemble closely the more conventional analytic properties of relativistic 1+1 dimensional integrable models~\cite{Zamolodchikov:1978xm}. It seems remarkable to us that it is possible to combine these into one coupled integrable system. 

Integrability of string theory investigated in the present paper is preserved under a number of  deformations. One can turn on moduli~\cite{OhlssonSax:2018hgc} or NS-NS flux~\cite{Hoare:2013pma,Hoare:2013ida,Hoare:2013lja,Lloyd:2014bsa} or consider the family of $\AdS_3\times\Sphere^3\times\Sphere^3\times\Sphere^1$ geometries~\cite{Babichenko:2009dk,OhlssonSax:2011ms,Sundin:2012gc,Borsato:2015mma} all of which are exact integrable string theory backgrounds. We therefore expect these theories to also exhibit the striking interplay of non-local spin chain and integrable field theory analyticity that we found in this paper. We intend to examine this in the near future.

\section*{Acknowledgements}

We would like to thank Alessandro Torrielli for many valuable and insightful discussions and for detailed comments on the manuscript. We want to thank S\'ebastien Leurent, Marius de Leeuw and Dima Volin for discussions, and Kostya Zarembo for discussions and comments on the manuscript.  BS acknowledges funding support from an STFC Consolidated Grant ``Theoretical Physics at City University'' ST/P000797/1. BS is grateful for the hospitality of Perimeter Institute where part of this work was carried out. Research at Perimeter Institute is supported in part by the Government of Canada through the Department of Innovation, Science and Economic Development Canada and by the Province of Ontario through the Ministry of Economic Development, Job Creation and Trade. 
BS is greatful to the organisers of the ``Exact Computations AdS/CFT'' workshop at CERN for providing a stimulating atmosphere during parts of this project.

\appendix

\section{On minimal solutions of crossing}
\label{sec:naturalness}

Imposing unitarity and LR symmetry, the minimal normalisation of the $\AdS_3 \times \Sphere^3 \times \Torus^4$ S matrix is given by
\begin{equation}
  \begin{aligned}
    S^{\sL\sL}_0(p_1,p_2) &= \biggl( \frac{x_1^+}{x_1^-} \frac{x_2^-}{x_2^+} \biggr)^{1/2} \frac{x_1^- - x_2^+}{x_1^+ - x_2^-} \Sigma^{\sL\sL}_{p_1p_2} , \\
    S^{\sL\sR}_0(p_1,p_2) &= \biggl( \frac{x_1^-}{x_1^+} \frac{x_2^-}{x_2^+} \biggr)^{1/2} \frac{1 - \frac{1}{x_1^+ x_2^+}}{1 - \frac{1}{x_1^- x_2^-}} \Sigma^{\sL\sR}_{p_1p_2} , \\
    S^{\sR\sL}_0(p_1,p_2) &= \biggl( \frac{x_1^+}{x_1^-} \frac{x_2^+}{x_2^-} \biggr)^{1/2} \frac{1 - \frac{1}{x_1^- x_2^-}}{1 - \frac{1}{x_1^+ x_2^+}} \Sigma^{\sR\sL}_{p_1p_2} , \\
    S^{\sR\sR}_0(p_1,p_2) &= \biggl( \frac{x_1^-}{x_1^+} \frac{x_2^+}{x_2^-} \biggr)^{1/2} \frac{x_1^+ - x_2^-}{x_1^- - x_2^+} \Sigma^{\sR\sR}_{p_1p_2} .
    \label{eq:min-norm-choice}
  \end{aligned}
\end{equation}
The remaining phases $\Sigma^{IJ}_{p_1p_2}$ satisfy the crossing equations\footnote{
  These crossing equations are minimal in the sense that the right hand side is the square root of the right hand side of the double crossing equations, which means that the solution is the simplest solution that solves the double crossing equation.
}
\begin{equation}
  \Sigma^{\sL\sL}_{p_1p_2} \Sigma^{\sR\sL}_{\bar{p}_1p_2} = \frac{x_1^+ - x_2^+}{x_1^+ - x_2^-} \frac{x_1^- - x_2^-}{x_1^- - x_2^+} , \qquad
  \Sigma^{\sL\sL}_{\bar{p}_1p_2} \Sigma^{\sR\sL}_{p_1p_2} = \frac{1-\frac{1}{x_1^+x_2^-}}{1-\frac{1}{x_1^+x_2^+}} \frac{1-\frac{1}{x_1^-x_2^+}}{1-\frac{1}{x_1^-x_2^-}} ,
\end{equation}
which have solution of the form
\begin{equation}
  \Sigma^{\sL\sL}_{p_1p_2} = e^{-2i\theta^{\sL\sL}(x_1^{\pm},x_2^{\pm})} , \qquad
  \Sigma^{\sL\sR}_{p_1p_2} = e^{-2i\theta^{\sL\sR}(x_1^{\pm},x_2^{\pm})} ,
\end{equation}
with
\begin{equation}
  \chi^{\sL\sL}(x,y) = \frac{1}{2} \bigl( \chi^{\text{HL}}(x,y) + \chi^-(x,y) \bigr) , \qquad
  \chi^{\sL\sR}(x,y) = \frac{1}{2} \bigl( \chi^{\text{HL}}(x,y) - \chi^-(x,y) \bigr) .
  \label{eq:soln-hl-ord-exact}
\end{equation}
Following the analysis of~\cite{Borsato:2013hoa}, $\chi^{\sL\sL}(x,y)$ is regular both at $x=y$ and $x=1/y$, while $\chi^{\sL\sR}(x,y)$ has a simple zero at $x=1/y$. Taking this into account we see that $S^{\sL\sL}_0(p_1p_2)$ has the expected S channel pole at $x_1^+ = x_2^-$ and $S^{\sL\sR}_0(p_1,p_2)$ has the expected T channel pole at $x_1^+ = 1/x_2^-$.

However, we do not have any of the expected DHM type double poles. Hence we are led to search for a phase which provides those poles and which solves the homogeneous crossing equation. In the pure RR case, such a solution is given by
\begin{equation}
  \Sigma^{\text{hom}}_{p_1p_2} = \biggl( \frac{x_1^+}{x_1^-} \frac{x_2^-}{x_2^+} \biggr)^{1/2} \frac{1-\frac{1}{x_1^-x_2^+}}{1-\frac{1}{x_1^+x_2^-}} \sigma^{-2}_{\text{even}}(p_1,p_2) ,
  \label{eq:soln-hom}
\end{equation}
where
\begin{equation}
  \sigma_{\text{even}}(p_1,p_2) = \sigma_{\text{BES}}(p_1,p_2) \sigma^{-1}_{\text{HL}}(p_1,p_2)
\end{equation}
is the part of the BES phase that is invariant under double crossing. Once this homogeneous phase is included we arrive at the normalisation of~\cite{Borsato:2014hja}, which we use in the main text.

\section{Matrix representations}
\label{sec:matrix-representations}

For completeness, in this section we write down explicit two-particle representations in the massive LL and LR sectors of $\AdS_3$, and show how short subrepresentations corresponding to potential bound states appear. We follow the conventions of~\cite{Borsato:2014hja} and work in the spin-chain frame.

\subsection{Fundamental representations}

The L representation acts on the basis $\bigl( \ket{\phi^{\sL}_p}, \ket{\psi^{\sL}_p} \bigr)$, and the supercharges take the form
\begin{equation}
  \begin{aligned}
    \gen{Q}_{\sL} &= \eta_p \begin{pmatrix} 0 & 0 \\ 1 & 0 \end{pmatrix} , \qquad &
    \gen{Q}_{\sR} &= \frac{\eta_p}{x_p^-} \begin{pmatrix} 0 & -1 \\ 0 & 0 \end{pmatrix} , \\
    \bar{\gen{Q}}_{\sL} &= \eta_p \begin{pmatrix} 0 & 1 \\ 0 & 0 \end{pmatrix} , \qquad &
    \bar{\gen{Q}}_{\sR} &= \frac{\eta_p}{x_p^+} \begin{pmatrix} 0 & 0 \\ -1 & 0 \end{pmatrix} ,
  \end{aligned}
\end{equation}
where
\begin{equation}
  \eta_p = \sqrt{\frac{ih}{2} ( x_p^- - x_p^+ ) } .
\end{equation}
Similarly, the R representation acts on the basis $\bigl( \ket{\psi^{\sR}_p}, \ket{\phi^{\sR}_p} \bigr)$, and the supercharges take the form
\begin{equation}
  \begin{aligned}
    \gen{Q}_{\sL} &= \frac{\eta_p}{x_p^-} \begin{pmatrix} 0 & 0 \\ -1 & 0 \end{pmatrix} , \qquad &
    \gen{Q}_{\sR} &= \eta_p \begin{pmatrix} 0 & 1 \\ 0 & 0 \end{pmatrix} , \\
    \bar{\gen{Q}}_{\sL} &= \frac{\eta_p}{x_p^+} \begin{pmatrix} 0 & -1 \\ 0 & 0 \end{pmatrix} , \qquad &
    \bar{\gen{Q}}_{\sR} &= \eta_p \begin{pmatrix} 0 & 0 \\ 1 & 0 \end{pmatrix} .
  \end{aligned}  
\end{equation}
For real momentum $p$, both the L and R representations are unitary.

In the following subsections we will consider states created as tensor products of fundamental representations. It will then be important to take into account the non-trivial coproduct as discussed in~\cite{Borsato:2013qpa,Borsato:2014hja}.

\subsection{LL bound states}

Let us start working in the basis $\bigl( \ket{\phi^{\sL}_{p_1} \phi^{\sL}_{p_2}}, \ket{\phi^{\sL}_{p_1} \psi^{\sL}_{p_2}}, \ket{\psi^{\sL}_{p_1} \phi^{\sL}_{p_2}}, \ket{\psi^{\sL}_{p_1} \psi^{\sL}_{p_2}}\bigr)$, and apply a similarity transformation using the matrix
\begin{equation}
  U = \frac{1}{\sqrt{\eta_{p_1}^2 + \eta_{p_2}^2}}
  \begin{pmatrix}
    \sqrt{\eta_{p_1}^2 + \eta_{p_2}^2} & 0 & 0 & 0 \\
    0 & \sqrt{\frac{x_{p_1}^+}{x_{p_1}^-}} \eta_{p_2} & -\eta_{p_1} & 0 & 0 \\
    0 & \eta_{p_1} & \sqrt{\frac{x_{p_1}^-}{x_{p_1}^+}} \eta_{p_2} & 0 \\
    0 & 0 & 0 & \sqrt{\eta_{p_1}^2 + \eta_{p_2}^2}
  \end{pmatrix} .
\end{equation}
In the new basis the L supercharges are
\begin{equation}
  \gen{Q}_{\sL} = \sqrt{\eta_{p_1}^2 + \eta_{p_2}^2} \begin{pmatrix} 0&0&0&0\\ 1&0&0&0 \\ 0&0&0&0 \\ 0&0&-1&0 \end{pmatrix} ,
  \qquad
  \bar{\gen{Q}}_{\sL} = \sqrt{\eta_{p_1}^2 + \eta_{p_2}^2} \begin{pmatrix} 0&1&0&0\\ 0&0&0&0 \\ 0&0&0&-1 \\ 0&0&0&0 \end{pmatrix} ,
\end{equation}
and the R supercharges are
\begin{equation}
  \begin{aligned}
    \bar{\gen{Q}}_{\sR} &=
    \frac{\eta_{p_1}\eta_{p_2}}{x_{p_1}^+ x_{p_2}^+ \sqrt{\eta_{p_1}^2 + \eta_{p_2}^2}}
    \begin{pmatrix}
      0 & 0 & 0 & 0 \\
      - x_{p_2}^+ \frac{\eta_{p_1}}{\eta_{p_2}} - x_{p_1}^- \frac{\eta_{p_2}}{\eta_{p_1}} & 0 & 0 & 0 \\
      \sqrt{\frac{x_{p_1}^+}{x_{p_1}^-}} (x_{p_1}^- - x_{p_2}^+) & 0 & 0 & 0 \\
      0 & \sqrt{\frac{x_{p_1}^+}{x_{p_1}^-}} (x_{p_1}^- - x_{p_2}^+) & x_{p_2}^+ \frac{\eta_{p_1}}{\eta_{p_2}} + x_{p_1}^- \frac{\eta_{p_2}}{\eta_{p_1}} & 0
    \end{pmatrix} ,
    \\
    \gen{Q}_{\sR} &=
    \frac{\eta_{p_1} \eta_{p_2}}{x_{p_1}^- x_{p_2}^- \sqrt{\eta_{p_1}^2 + \eta_{p_2}^2}}
    \begin{pmatrix}
      0 & - x_{p_2}^- \frac{\eta_{p_1}}{\eta_{p_2}} - x_{p_1}^+ \frac{\eta_{p_2}}{\eta_{p_1}} & \sqrt{\frac{x_{p_1}^-}{x_{p_1}^+}} (x_{p_1}^+ - x_{p_2}^-)  & 0 \\
      0 & 0 & 0 & \sqrt{\frac{x_{p_1}^-}{x_{p_1}^+}} (x_{p_1}^+ - x_{p_2}^-) \\
      0 & 0 & 0 & x_{p_2}^- \frac{\eta_{p_1}}{\eta_{p_2}} + x_{p_1}^+ \frac{\eta_{p_2}}{\eta_{p_1}} \\
      0 & 0 & 0 & 0
    \end{pmatrix} .
  \end{aligned}
\end{equation}
For real $p_1$ and $p_2$ we have $(x_{p_1}^+)^* = x_{p_1}^-$ and $(x_{p_2}^+)^* = x_{p_2}^-$, so that $U^{\dag} = U$ is unitary and $\gen{Q}_{\sL}^{\dag} = \bar{\gen{Q}}_{\sL}$ and $\gen{Q}_{\sR}^{\dag} = \bar{\gen{Q}}_{\sR}$.

\paragraph{Symmetric BPS state.}

Let us now set
\begin{equation}
  x_{p_1}^+ = X_p^+, \qquad x_{p_2}^- = X_p^- , \qquad x_{p_1}^- = x_{p_2}^+ = X_p .
\end{equation}
The supercharges are then
\begin{equation}
  \gen{Q}_{\sL} = \eta_p \begin{pmatrix} 0&0&0&0\\ 1&0&0&0 \\ 0&0&0&0 \\ 0&0&-1&0 \end{pmatrix} ,
  \quad
  \bar{\gen{Q}}_{\sL} = \eta_p \begin{pmatrix} 0&1&0&0\\ 0&0&0&0 \\ 0&0&0&-1 \\ 0&0&0&0 \end{pmatrix} ,
  \quad
  \bar{\gen{Q}}_{\sR} = \frac{\eta_p}{X_p^+} \begin{pmatrix} 0&0&0&0\\ 1&0&0&0 \\ 0&0&0&0 \\ 0&0&-1&0 \end{pmatrix} ,
\end{equation}
and
\begin{equation}
  \gen{Q}_{\sR} =
  -\frac{\eta_p}{X_p^-}
  \begin{pmatrix}
    0 & 1 & \sqrt{\bigl(\frac{1}{X_p^+} - \frac{1}{X_p}\bigr)( X_p^- - X_p )} & 0 \\
    0 & 0 & 0 & \sqrt{\bigl(\frac{1}{X_p^+} - \frac{1}{X_p}\bigr)( X_p^- - X_p )} \\
    0 & 0 & 0 & -1 \\
    0 & 0 & 0 & 0
  \end{pmatrix} .
\end{equation}
Since $(x_{p_1}^+)^*$ and $(x_{p_2}^+)^*$  are no longer equal to $x_{p_1}^-$ and $x_{p_2}^-$, we no longer have that $\gen{Q}_{\sR}^{\dag}$ equals $\bar{\gen{Q}}_{\sR}$. However, if we consider the closed subalgebra generated by the upper left $2\times2$ block of each matrix we have
\begin{equation}
  \begin{aligned}
    \gen{Q}_{\sL} &\to \eta_p \begin{pmatrix} 0&0\\1&0 \end{pmatrix} , \qquad &
    \gen{Q}_{\sR} &\to \frac{\eta_p}{X_p^-} \begin{pmatrix} 0&-1\\0&0 \end{pmatrix} , \\
    \bar{\gen{Q}}_{\sL} &\to \eta_p \begin{pmatrix} 0&1\\0&0 \end{pmatrix} , &
    \bar{\gen{Q}}_{\sR} &\to \frac{\eta_p}{X_p^+} \begin{pmatrix} 0&0\\-1&0 \end{pmatrix} .
  \end{aligned}
\end{equation}
This is just a short representation with momentum $p$.

\paragraph{Anti-symmetric BPS state.}

If we instead set
\begin{equation}
  x_{p_2}^+ = X_p^+, \qquad x_{p_1}^- = X_p^- , \qquad x_{p_1}^+ = x_{p_2}^- = X_p
\end{equation}
we find
\begin{equation}
  \gen{Q}_{\sL} = \eta_p \begin{pmatrix} 0&0&0&0\\ 1&0&0&0 \\ 0&0&0&0 \\ 0&0&-1&0 \end{pmatrix} ,
  \quad
  \bar{\gen{Q}}_{\sL} = \eta_p \begin{pmatrix} 0&1&0&0\\ 0&0&0&0 \\ 0&0&0&-1 \\ 0&0&0&0 \end{pmatrix} ,
  \quad
  \gen{Q}_{\sR} = \frac{\eta_p}{X_p^-} \begin{pmatrix} 0&-1&0&0\\ 0&0&0&0 \\ 0&0&0&1 \\ 0&0&0&0 \end{pmatrix} ,
\end{equation}
and
\begin{equation}
  \bar{\gen{Q}}_{\sR} =
  \frac{\eta_p}{X_p^+}
  \begin{pmatrix}
    0 & 0 & 0 & 0 \\
    1 & 0 & 0 & 0 \\
    \sqrt{\bigl(\frac{1}{X_p^-} - \frac{1}{X_p}\bigr)( X_p^+ - X_p )} & 0 & 0 & 0 \\
    0 & \sqrt{\bigl(\frac{1}{X_p^-} - \frac{1}{X_p}\bigr)( X_p^+ - X_p )} & -1 & 0
  \end{pmatrix} .
\end{equation}
In this case the invariant submodule sits in the lower right corner where the charges are give by
\begin{equation}
  \begin{aligned}
    \gen{Q}_{\sL} &\to \eta_p \begin{pmatrix} 0&0\\-1&0 \end{pmatrix} , \qquad &
    \gen{Q}_{\sR} &\to \frac{\eta_p}{X_p^-} \begin{pmatrix} 0&1\\0&0 \end{pmatrix} , \\
    \bar{\gen{Q}}_{\sL} &\to \eta_p \begin{pmatrix} 0&-1\\0&0 \end{pmatrix} , &
    \bar{\gen{Q}}_{\sR} &\to \frac{\eta_p}{X_p^+} \begin{pmatrix} 0&0\\1&0 \end{pmatrix} ,
  \end{aligned}
\end{equation}
which, up to an irrelevant overall sign, is the same two-dimensional L representation we saw above.

\subsection{LR bound states}
\label{app:lr-bound-states}

Let us now consider the LR tensor product. We start with the basis ( $\ket{\phi^{\sL}_{p_1} \phi^{\sR}_{p_2}}$, $\ket{\phi^{\sL}_{p_1} \psi^{\sR}_{p_2}}$, $\ket{\psi^{\sL}_{p_1} \phi^{\sR}_{p_2}}$, $\ket{\psi^{\sL}_{p_1} \psi^{\sR}_{p_2}}$), and apply a similarity transformation using
\begin{equation}
  U =
  \frac{1}{\sqrt{\eta_{p_1}^2 + \frac{\eta_{p_2}^2}{x_{p_2}^+x_{p_2}^-}}}
  \begin{pmatrix}
    \sqrt{\eta_{p_1}^2 + \frac{\eta_{p_2}^2}{x_{p_2}^+x_{p_2}^-}} & 0 & 0 & 0 \\
    0 & -\sqrt{\frac{x_{p_1}^-}{x_{p_1}^+}} \frac{\eta_{p_2}}{x_{p_2}^+} & - \eta_{p_1} & 0 \\
    0 & \eta_{p_1} & -\sqrt{\frac{x_{p_1}^-}{x_{p_1}^+}} \frac{\eta_{p_2}}{x_{p_2}^-} & 0 \\
    0 & 0 & 0 & \sqrt{\eta_{p_1}^2 + \frac{\eta_{p_2}^2}{x_{p_2}^+x_{p_2}^-}} .
  \end{pmatrix}
\end{equation}
Note that $U$ is unitary for real $p_1$ and $p_2$. The supercharges in the new basis are
\begin{equation}
  \gen{Q}_{\sL} =
  \sqrt{\eta_{p_1}^2 + \frac{\eta_{p_2}^2}{x_{p_2}^+x_{p_2}^-}}
  \begin{pmatrix}
    0 & 1 & 0 & 0 \\
    0 & 0 & 0 & 0 \\
    0 & 0 & 0 &-1 \\
    0 & 0 & 0 & 0
  \end{pmatrix} ,
  \quad
  \bar{\gen{Q}}_{\sL} =
  \sqrt{\eta_{p_1}^2 + \frac{\eta_{p_2}^2}{x_{p_2}^+x_{p_2}^-}}
  \begin{pmatrix}
    0 & 0 & 0 & 0 \\
    1 & 0 & 0 & 0 \\
    0 & 0 & 0 & 0 \\
    0 & 0 &-1 & 0
  \end{pmatrix} ,
\end{equation}
and
\begin{equation}
  \begin{aligned}
    \bar{\gen{Q}}_{\sR} &=
    \frac{\eta_{p_1} \eta_{p_2}}{\sqrt{\eta_{p_1}^2 + \frac{\eta_{p_2}^2}{x_{p_2}^+x_{p_2}^-}}}
    \begin{pmatrix}
      0 & \frac{1}{x_{p_1}^+}\frac{\eta_{p_1}}{\eta_{p_2}} + \frac{x_{p_1}^-}{x_{p_1}^+} \frac{1}{x_{p_2}^+} \frac{\eta_{p_2}}{\eta_{p_1}} & \sqrt{\frac{x_{p_1}^-}{x_{p_1}^+}} \bigl(1 - \frac{1}{x_{p_1}^- x_{p_2}^-}\bigr) & 0 \\
      0 & 0 & 0 & \sqrt{\frac{x_{p_1}^-}{x_{p_1}^+}} \bigl(1 - \frac{1}{x_{p_1}^- x_{p_2}^-}\bigr) \\
      0 & 0 & 0 & -\frac{1}{x_{p_1}^+} \frac{\eta_{p_1}}{\eta_{p_2}} - \frac{x_{p_1}^-}{x_{p_1}^+} \frac{1}{x_{p_2}^+} \frac{\eta_{p_2}}{\eta_{p_1}} \\
      0 & 0 & 0 & 0
    \end{pmatrix} ,
    \\
    \gen{Q}_{\sR} &=
    \frac{\eta_{p_1} \eta_{p_2}}{\sqrt{\eta_{p_1}^2 + \frac{\eta_{p_2}^2}{x_{p_2}^+x_{p_2}^-}}}
    \begin{pmatrix}
      0 &  0 & 0 & 0 \\
      \frac{1}{x_{p_1}^-} \frac{\eta_{p_1}}{\eta_{p_2}} + \frac{x_{p_1}^+}{x_{p_1}^-} \frac{1}{x_{p_2}^-} \frac{\eta_{p_2}}{\eta_{p_1}} & 0 & 0 & 0 \\
      \sqrt{\frac{x_{p_1}^+}{x_{p_1}^-}} \bigl(1 - \frac{1}{x_{p_1}^+ x_{p_2}^+}\bigr) & 0 & 0 & 0 \\
      0 & \sqrt{\frac{x_{p_1}^+}{x_{p_1}^-}} \bigl(1 - \frac{1}{x_{p_1}^+ x_{p_2}^+}\bigr) & -\frac{1}{x_{p_1}^-} \frac{\eta_{p_1}}{\eta_{p_2}} - \frac{x_{p_1}^+}{x_{p_1}^-} \frac{1}{x_{p_2}^-} \frac{\eta_{p_2}}{\eta_{p_1}} & 0
    \end{pmatrix} .
  \end{aligned}
\end{equation}
Again this representation is unitary for real $p_1$ and $p_2$.

\paragraph{Symmetric BPS state.}

Let us set
\begin{equation}\label{eq:lr-l-bps-condition}
  x_{p_1}^+ = \frac{1}{x_{p_2}^+} = X_p , \qquad
  x_{p_1}^- = X_p^- , \qquad
  x_{p_2}^- = \frac{1}{X_p^+} .
\end{equation}
We get
\begin{equation}
  \gen{Q}_{\sL} =
  \eta_p
  \begin{pmatrix}
    0 & 1 & 0 & 0 \\
    0 & 0 & 0 & 0 \\
    0 & 0 & 0 &-1 \\
    0 & 0 & 0 & 0
  \end{pmatrix} ,
  \quad
  \bar{\gen{Q}}_{\sL} =
  \eta_p
  \begin{pmatrix}
    0 & 0 & 0 & 0 \\
    1 & 0 & 0 & 0 \\
    0 & 0 & 0 & 0 \\
    0 & 0 &-1 & 0
  \end{pmatrix} ,
  \quad
  \gen{Q}_{\sR} =
  \frac{\eta_p}{X_p^-}
  \begin{pmatrix}
    0 & 0 & 0 & 0 \\
    -1 & 0 & 0 & 0 \\
    0 & 0 & 0 & 0 \\
    0 & 0 & 1 & 0
  \end{pmatrix} ,
\end{equation}
and
\begin{equation}
  \bar{\gen{Q}}_{\sR} =
  \eta_p
  \begin{pmatrix}
    0 & -\frac{1}{X_p^+} & \sqrt{\Bigl(\frac{1}{X_p^+}-\frac{1}{X}\Bigr)\Bigl(\frac{1}{X}-\frac{1}{X_p^-}\Bigr)} & 0 \\
    0 & 0 & 0 & \sqrt{\Bigl(\frac{1}{X_p^+}-\frac{1}{X}\Bigr)\Bigl(\frac{1}{X}-\frac{1}{X_p^-}\Bigr)} \\
    0 & 0 & 0 & \frac{1}{X_p^+} \\
    0 & 0 & 0 & 0
  \end{pmatrix} .
\end{equation}
The upper left block reduces to a short fundamental L representation.

Instead of the condition~\eqref{eq:lr-l-bps-condition} we can also set
\begin{equation}\label{eq:lr-r-bps-condition}
  x_{p_1}^+ = \frac{1}{x_{p_2}^+} = \frac{1}{X_p} , \qquad
  x_{p_1}^- = \frac{1}{X_p^+} , \qquad
  x_{p_2}^- = X_p^- ,
\end{equation}
which results in a short R sub-representation. Formally, the conditions~\eqref{eq:lr-l-bps-condition} and~\eqref{eq:lr-r-bps-condition} are related by a simple relabelling $X_p^{\pm} \to 1/X_p^{\mp}$, $X_p \to 1/X_p$. However, the physical interpretation is different. In particular, if we assume that the bound state lives in the physical region $|X_p^{\pm}| > 1$, then the L sub-representation has $|x_{p_1}^{\pm}| > 1$ and $|x_{p_2}^{\pm}| < 1$, while the R sub-representation has $|x_{p_1}^{\pm}| < 1$ and $|x_{p_2}^{\pm}| > 1$.

\paragraph{Anti-symmetric BPS state.}

If we instead set
\begin{equation}
  x_{p_1}^- = \frac{1}{x_{p_2}^-} = X_p , \qquad
  x_{p_1}^+ = X_p^+ , \qquad
  x_{p_2}^+ = \frac{1}{X_p^-} .
\end{equation}
We get
\begin{equation}\nonumber
  \gen{Q}_{\sL} =
  \eta_p
  \begin{pmatrix}
    0 & 1 & 0 & 0 \\
    0 & 0 & 0 & 0 \\
    0 & 0 & 0 &-1 \\
    0 & 0 & 0 & 0
  \end{pmatrix} ,
  \quad
  \bar{\gen{Q}}_{\sL} =
  \eta_p
  \begin{pmatrix}
    0 & 0 & 0 & 0 \\
    1 & 0 & 0 & 0 \\
    0 & 0 & 0 & 0 \\
    0 & 0 &-1 & 0
  \end{pmatrix} ,
  \quad
  \bar{\gen{Q}}_{\sR} =
  \frac{\eta_p}{X_p^+}
  \begin{pmatrix}
    0 &-1 & 0 & 0 \\
    0 & 0 & 0 & 0 \\
    0 & 0 & 0 & 1 \\
    0 & 0 & 0 & 0
  \end{pmatrix} ,
\end{equation}
and
\begin{equation}
  \gen{Q}_{\sR} =
  \eta_p
  \begin{pmatrix}
      0 & 0 & 0 & 0 \\
      -\frac{1}{X_p^-} & 0 & 0 & 0 \\
      \sqrt{\Bigl(\frac{1}{X_p^+}-\frac{1}{X}\Bigr)\Bigl(\frac{1}{X}-\frac{1}{X_p^-}\Bigr)} & 0 & 0 & 0 \\
      0 & \sqrt{\Bigl(\frac{1}{X_p^+}-\frac{1}{X}\Bigr)\Bigl(\frac{1}{X}-\frac{1}{X_p^-}\Bigr)} & \frac{1}{X_p^-} & 0
  \end{pmatrix} .
\end{equation}
The lower right block reduces to a fundamental L sub-representation. As above, we find an R representation instead if we send $X_p^{\pm} \to 1/X_p^{\mp}$ and $X_p \to 1/X_p$.

\section{Spin-chain bound states}
\label{sec:spin-chain-bound-states}

As a toy model for bound states let us consider a one-dimensional infinite lattice, with a nilpotent creation operator $\gen{J}^{\dag}_n$ which creates an excitation on site $n$, and short-range interactions. A wave function describing two excitations with momentum $p_1$ and $p_2$, with $p_1 > p_2$ takes the form
\begin{equation}
  \Psi(p_1,p_2) = A(p_1,p_2) \sum_{n_1<n_2} \Bigl( e^{i(p_1 n_1 + p_2 n_2)} + S(p_1,p_2) \, e^{i(p_1 n_2 + p_2 n_1)} \Bigr) \gen{J}^{\dag}_{n_1} \gen{J}^{\dag}_{n_2} .
\end{equation}
In the above expression, the first term describes an incoming wave and the second term an outgoing wave. $A(p_1,p_2)$ is a normalisation factor and $S(p_1,p_2)$ gives the scattering phase. We now continue the momenta to complex values and introduce
\begin{equation}
  p_1 = \frac{p}{2} + iq , \qquad
  p_2 = \frac{p}{2} - iq , \qquad
  m = \frac{n_2+n_1}{2} , \qquad
  r = \frac{n_2-n_1}{2} .
\end{equation}
Since $n_1 < n_2$ we have that $r > 0$. The wave function can now be written as\footnote{To be fully correct one needs to be careful about the summation ranges in the following expression. However these subtleties play no role in the simple discussion here.}
\begin{equation}
  \Psi(p_1,p_2) = A(p_1,p_2) \sum_{m,r} e^{ipm} \Bigl( e^{2qr} + S(p_1,p_2) e^{-2qr} \Bigr) \gen{J}^{\dag}_{n_1} \gen{J}^{\dag}_{n_2} .
\end{equation}
If the real part of $q$ is non-zero, either the first or second term in the above sum diverges for large $r$.
\begin{itemize}
\item If $\Re q > 0$, the incoming wave is divergent. In order to have a normalisable wave function we need to set $A(p_1,p_2) = 0$. However, this leads to a vanishing wave function unless $S(p_1,p_2)$ has a pole at this location.
\item If $\Re q < 0$, the outgoing wave is divergent, and the wave function can only be normalisable if $S(p_1,p_2) = 0$.
\end{itemize}
In either case, the resulting bound state wave function takes exactly the same form,\footnote{Switching the sign of $q$ exchanges $p_1$ and $p_2$ and hence the incoming and outgoing part of the wave function. In this simple example we consider two identical excitations, which means that the two bound states are identical. This would not necessarily be the case if we considered a model with excitations carrying additional quantum numbers.} and describes a localised wave packet moving with momentum $p$.

From the above discussion we conclude that a simple pole of the scattering phase can correspond to a physical bound state only if the imaginary part of the momentum of the first particle is positive.

Note that the picture of a localised wave packet consisting of two excitations travelling with the same momentum is simplest when $p$ and $q$ are both real. Let us consider the two-particle bound state in the LL sector of $\AdS_3$ discussed in the main text. The two excitations are described by $x^{\pm}$ and $y^{\pm}$, satisfying
\begin{equation}
  x^+ + \frac{1}{x^+} - x^- - \frac{1}{x^-} =
  \frac{2i}{h} = 
  y^+ + \frac{1}{y^+} - y^- - \frac{1}{y^-}
\end{equation}
and
\begin{equation}
  x^+ = y^- , \qquad z^+ = y^+ , \qquad z^- = x^- ,
\end{equation}
where we introduced the parameters $z^{\pm}$ to describe the bound state.
For real bound state momentum $p$ we can parametrise the bound state by
\begin{equation}
  z^{\pm} = \frac{2 + \sqrt{4 + 4h^2\sin^2\frac{p}{2}}}{2h\sin\frac{p}{2}} e^{\pm\frac{ip}{2}} .
\end{equation}
Writing the momenta of the two fundamental excitations as
\begin{equation}
  p_1 = -i\log\frac{x^+}{x^-} = \frac{p}{2} + i q , \qquad
  p_2 = -i\log\frac{y^+}{y^-} = \frac{p}{2} - i q ,
\end{equation}
we find that $q$ is real for
\begin{equation}
  h < \frac{\cos\frac{p}{2}}{\sin^2\frac{p}{2}} .
\end{equation}
As long as the coupling is small enough, $q$ is purely real and all parameters $x^+$ (and thus $y^-$) sits on the real line outside the unit circle. However, if we keep the momentum $p$ fixed and increase the coupling constant, $x^+$ at some point hits the unit circle and move off the real line, which means that $q$ acquires an imaginary part.\footnote{For a discussion of this phenomena in the context of string theory on $\AdS_5 \times \Sphere^5$ see reference~\cite{Arutyunov:2007tc}} When this happens the bound state momentum is no longer evenly divided between the two fundamental excitations. However, the bound state condition discussed above still holds.

\bibliographystyle{nb}
\bibliography{refs}

\makeatletter \@ifundefined{Sphere}{\newcommand{\Sphere}{S}}{}
  \@ifundefined{AdS}{\newcommand{\AdS}{AdS}}{}
  \@ifundefined{CFT}{\newcommand{\CFT}{\text{CFT}}}{}
  \@ifundefined{CP}{\newcommand{\CP}{CP}}{}
  \@ifundefined{Torus}{\newcommand{\Torus}{T}}{}
  \@ifundefined{superN}{\newcommand{\superN}{\mathcal{N}}}{}
  \@ifundefined{grpOSp}{\newcommand{\grpOSp}{\text{OSp}}}{}
  \@ifundefined{grpPSU}{\newcommand{\grpPSU}{\text{PSU}}}{}
  \@ifundefined{grpSU}{\newcommand{\grpSU}{\text{SU}}}{}
  \@ifundefined{grpU}{\newcommand{\grpU}{\text{U}}}{}
  \@ifundefined{grpD}{\newcommand{\grpD}{\text{D}}}{}
  \@ifundefined{grpSL}{\newcommand{\grpSL}{\text{SL}}}{}
  \@ifundefined{grpSp}{\newcommand{\grpSp}{\text{Sp}}}{}
  \@ifundefined{grpUSp}{\newcommand{\grpUSp}{\text{USp}}}{}
  \@ifundefined{grpSO}{\newcommand{\grpSO}{\text{SO}}}{}
  \@ifundefined{grpO}{\newcommand{\grpO}{\text{O}}}{}
  \@ifundefined{algOSp}{\newcommand{\algOSp}{\text{osp}}}{}
  \@ifundefined{algPSU}{\newcommand{\algPSU}{\text{psu}}}{}
  \@ifundefined{algSU}{\newcommand{\algSU}{\text{su}}}{}
  \@ifundefined{algSp}{\newcommand{\algSp}{\text{sp}}}{}
  \@ifundefined{algSL}{\newcommand{\algSL}{\text{sl}}}{}
  \@ifundefined{algGL}{\newcommand{\algGL}{\text{gl}}}{}
  \@ifundefined{algU}{\newcommand{\algU}{\text{u}}}{}
  \@ifundefined{algSO}{\newcommand{\algSO}{\text{so}}}{}
  \@ifundefined{algO}{\newcommand{\algO}{\text{o}}}{}
  \@ifundefined{Integers}{\newcommand{\Integers}{\text{Z}}}{}
  \@ifundefined{Reals}{\newcommand{\Reals}{\text{R}}}{} \makeatother
%bibliography generated by nb.bst v1.01 (C) 2003-2010 Niklas Beisert
\begin{thebibliography}{10}
\ifx\href\asklfhas\newcommand{\href}[2]{#2}\fi
\ifx\arxivref\asklfhas\newcommand{\arxivref}[2]{\href{http://arxiv.org/abs/#1}{#2}}\fi
\ifx\doiref\asklfhas\newcommand{\doiref}[2]{\href{http://dx.doi.org/#1}{#2}}\fi
\raggedright
\small
\parskip 0pt

\bibitem{Maldacena:2001km}
J.~M.~Maldacena and H.~Ooguri,
\textit{``Strings in {$\AdS_3$} and the {$\grpSL(2,R)$} {WZW} model.~{III}:
  Correlation functions''},
\textsf{\doiref{10.1103/PhysRevD.65.106006}{Phys.~Rev.~D65,~106006~(2002)}},
\texttt{\arxivref{hep-th/0111180}{hep-th/0111180}}.
%%CITATION = HEP-TH/0111180;%%

\bibitem{Metsaev:2001bj}
R.~Metsaev,
\textit{``Type {IIB} {G}reen-{S}chwarz superstring in plane wave
  {R}amond-{R}amond background''},
\textsf{\doiref{10.1016/S0550-3213(02)00003-2}{Nucl.~Phys.~B625,~70~(2002)}},
\texttt{\arxivref{hep-th/0112044}{hep-th/0112044}}.
%%CITATION = HEP-TH/0112044;%%

\bibitem{Berenstein:2002jq}
D.~E.~Berenstein, J.~M.~Maldacena and H.~S.~Nastase,
\textit{``Strings in flat space and {pp} waves from {$\superN = 4$} super
  {Y}ang {M}ills''},
\textsf{\doiref{10.1088/1126-6708/2002/04/013}{JHEP~0204,~013~(2002)}},
\texttt{\arxivref{hep-th/0202021}{hep-th/0202021}}.
%%CITATION = HEP-TH/0202021;%%

\bibitem{Metsaev:2002re}
R.~Metsaev and A.~A.~Tseytlin,
\textit{``Exactly solvable model of superstring in {R}amond-{R}amond plane wave
  background''},
\textsf{\doiref{10.1103/PhysRevD.65.126004}{Phys.~Rev.~D65,~126004~(2002)}},
\texttt{\arxivref{hep-th/0202109}{hep-th/0202109}}.
%%CITATION = HEP-TH/0202109;%%

\bibitem{Arutyunov:2009ga}
G.~Arutyunov and S.~Frolov,
\textit{``Foundations of the {$\AdS_5 \times \Sphere^5$} Superstring. {P}art
  {I}''},
\textsf{\doiref{10.1088/1751-8113/42/25/254003}{J.~Phys.~A~A42,~254003~(2009)}},
\texttt{\arxivref{0901.4937}{arxiv:0901.4937}}.
%%CITATION = ARXIV:0901.4937;%%

\bibitem{Beisert:2010jr}
N.~Beisert et~al.,
\textit{``Review of {AdS/CFT} Integrability: An Overview''},
\textsf{\doiref{10.1007/s11005-011-0529-2}{Lett.~Math.~Phys.~99,~3~(2012)}},
\texttt{\arxivref{1012.3982}{arxiv:1012.3982}}.
%%CITATION = 1012.3982;%%

\bibitem{Maldacena:1997re}
J.~M.~Maldacena,
\textit{``The large {N} limit of superconformal field theories and
  supergravity''},
\textsf{Adv.~Theor.~Math.~Phys.~2,~231~(1998)},
\texttt{\arxivref{hep-th/9711200}{hep-th/9711200}}.
%%CITATION = HEP-TH/9711200;%%

\bibitem{Janik:2006dc}
R.~A.~Janik,
\textit{``The {$\AdS_5 \times \Sphere^5$} superstring worldsheet {S}-matrix and
  crossing symmetry''},
\textsf{\doiref{10.1103/PhysRevD.73.086006}{Phys.~Rev.~D73,~086006~(2006)}},
\texttt{\arxivref{hep-th/0603038}{hep-th/0603038}}.
%%CITATION = HEP-TH/0603038;%%

\bibitem{Castillejo:1955ed}
L.~Castillejo, R.~H.~Dalitz and F.~J.~Dyson,
\textit{``Low's scattering equation for the charged and neutral scalar
  theories''},
\textsf{\doiref{10.1103/PhysRev.101.453}{Phys.~Rev.~101,~453~(1956)}}.
%%CITATION = PHRVA,101,453;%%

\bibitem{Dorey:2006dq}
N.~Dorey,
\textit{``Magnon bound states and the {AdS/CFT} correspondence''},
\textsf{J.~Phys.~A39,~13119~(2006)},
\texttt{\arxivref{hep-th/0604175}{hep-th/0604175}}.
%%CITATION = HEP-TH/0604175;%%

\bibitem{Dorey:2007xn}
N.~Dorey, D.~M.~Hofman and J.~M.~Maldacena,
\textit{``On the singularities of the magnon {S}-matrix''},
\textsf{\doiref{10.1103/PhysRevD.76.025011}{Phys.~Rev.~D76,~025011~(2007)}},
\texttt{\arxivref{hep-th/0703104}{hep-th/0703104}}.
%%CITATION = HEP-TH/0703104;%%

\bibitem{Beisert:2006ib}
N.~Beisert, R.~Hern{\'a}ndez and E.~L{\'o}pez,
\textit{``A crossing-symmetric phase for {$\AdS_5 \times \Sphere^5$}''},
\textsf{\doiref{10.1088/1126-6708/2006/11/070}{JHEP~0611,~070~(2006)}},
\texttt{\arxivref{hep-th/0609044}{hep-th/0609044}}.
%%CITATION = HEP-TH/0609044;%%

\bibitem{Beisert:2006ez}
N.~Beisert, B.~Eden and M.~Staudacher,
\textit{``Transcendentality and crossing''},
\textsf{\doiref{10.1088/1742-5468/2007/01/P01021}{J.~Stat.~Mech.~0701,~P01021~(2007)}},
\texttt{\arxivref{hep-th/0610251}{hep-th/0610251}}.
%%CITATION = HEP-TH/0610251;%%

\bibitem{Babichenko:2009dk}
A.~Babichenko, B.~Stefa{\'n}ski,~jr. and K.~Zarembo,
\textit{``Integrability and the {$\AdS_3/\CFT_2$} correspondence''},
\textsf{\doiref{10.1007/JHEP03(2010)058}{JHEP~1003,~058~(2010)}},
\texttt{\arxivref{0912.1723}{arxiv:0912.1723}}.
%%CITATION = 0912.1723;%%

\bibitem{Cagnazzo:2012se}
A.~Cagnazzo and K.~Zarembo,
\textit{``{B}-field in {$\AdS_3/\CFT_2$} Correspondence and Integrability''},
\textsf{\doiref{10.1007/JHEP11(2012)133,
  10.1007/JHEP04(2013)003}{JHEP~1211,~133~(2012)}},
\texttt{\arxivref{1209.4049}{arxiv:1209.4049}}.
%%CITATION = ARXIV:1209.4049;%%

\bibitem{Borsato:2014exa}
R.~Borsato, O.~Ohlsson~Sax, A.~Sfondrini and B.~Stefa{\'n}ski,~jr.,
\textit{``Towards the all-loop worldsheet {S} matrix for {$\AdS_3 \times
  \Sphere^3 \times \Torus^4$}''},
\textsf{\doiref{10.1103/PhysRevLett.113.131601}{Phys.~Rev.~Lett.~113,~131601~(2014)}},
\texttt{\arxivref{1403.4543}{arxiv:1403.4543}}.
%%CITATION = ARXIV:1403.4543;%%

\bibitem{Borsato:2014hja}
R.~Borsato, O.~Ohlsson~Sax, A.~Sfondrini and B.~Stefa{\'n}ski,~jr,
\textit{``The complete {$\AdS_3 \times \Sphere^3 \times \Torus^4$} worldsheet
  {S}-matrix''},
\textsf{\doiref{10.1007/JHEP10(2014)066}{JHEP~1410,~66~(2014)}},
\texttt{\arxivref{1406.0453}{arxiv:1406.0453}}.
%%CITATION = ARXIV:1406.0453;%%

\bibitem{Lloyd:2014bsa}
T.~Lloyd, O.~Ohlsson~Sax, A.~Sfondrini and B.~Stefa{\'n}ski,~jr.,
\textit{``The complete worldsheet {S} matrix of superstrings on {$\AdS_3 \times
  \Sphere^3 \times \Torus^4$} with mixed three-form flux''},
\textsf{\doiref{10.1016/j.nuclphysb.2014.12.019}{Nucl.~Phys.~B891,~570~(2015)}},
\texttt{\arxivref{1410.0866}{arxiv:1410.0866}}.
%%CITATION = ARXIV:1410.0866;%%

\bibitem{Borsato:2013hoa}
R.~Borsato, O.~Ohlsson~Sax, A.~Sfondrini, B.~Stefa{\'n}ski,~jr. and
  A.~Torrielli,
\textit{``Dressing phases of {$\AdS_3/\CFT_2$}''},
\textsf{\doiref{10.1103/PhysRevD.88.066004}{Phys.~Rev.~D88,~066004~(2013)}},
\texttt{\arxivref{1306.2512}{arxiv:1306.2512}}.
%%CITATION = ARXIV:1306.2512;%%

\bibitem{Borsato:2016kbm}
R.~Borsato, O.~Ohlsson~Sax, A.~Sfondrini and B.~Stefa{\'n}ski,~jr.,
\textit{``On the spectrum of {$\AdS_3 \times \Sphere^3 \times \Torus^4$}
  strings with {R}amond-{R}amond flux''},
\textsf{\doiref{10.1088/1751-8113/49/41/41LT03}{J.~Phys.~A49,~41LT03~(2016)}},
\texttt{\arxivref{1605.00518}{arxiv:1605.00518}}.
%%CITATION = ARXIV:1605.00518;%%

\bibitem{Borsato:2016xns}
R.~Borsato, O.~Ohlsson~Sax, A.~Sfondrini, B.~Stefa{\'n}ski,~jr. and
  A.~Torrielli,
\textit{``On the Dressing Factors, {B}ethe Equations and {Y}angian Symmetry of
  Strings on {$\AdS_3 \times \Sphere^3 \times \Torus^4$}''},
\textsf{\doiref{10.1088/1751-8121/50/2/024004}{J.~Phys.~A50,~024004~(2017)}},
\texttt{\arxivref{1607.00914}{arxiv:1607.00914}}.
%%CITATION = ARXIV:1607.00914;%%

\bibitem{Borsato:2013qpa}
R.~Borsato, O.~Ohlsson~Sax, A.~Sfondrini, B.~Stefa{\'n}ski,~jr. and
  A.~Torrielli,
\textit{``The all-loop integrable spin-chain for strings on {$\AdS_3 \times
  \Sphere^3 \times \Torus^4$}: the massive sector''},
\textsf{\doiref{10.1007/JHEP08(2013)043}{JHEP~1308,~043~(2013)}},
\texttt{\arxivref{1303.5995}{arxiv:1303.5995}}.
%%CITATION = ARXIV:1303.5995;%%

\bibitem{Arutyunov:2004vx}
G.~Arutyunov, S.~Frolov and M.~Staudacher,
\textit{``{B}ethe ansatz for quantum strings''},
\textsf{\doiref{10.1088/1126-6708/2004/10/016}{JHEP~0410,~016~(2004)}},
\texttt{\arxivref{hep-th/0406256}{hep-th/0406256}}.
%%CITATION = HEP-TH/0406256;%%

\bibitem{Hernandez:2006tk}
R.~Hern{\'a}ndez and E.~L{\'o}pez,
\textit{``Quantum corrections to the string {B}ethe ansatz''},
\textsf{\doiref{10.1088/1126-6708/2006/07/004}{JHEP~0607,~004~(2006)}},
\texttt{\arxivref{hep-th/0603204}{hep-th/0603204}}.
%%CITATION = HEP-TH/0603204;%%

\bibitem{Vieira:2010kb}
P.~Vieira and D.~Volin,
\textit{``Review of {AdS/CFT} Integrability, {C}hapter {III.3}: The dressing
  factor''},
\textsf{\doiref{10.1007/s11005-011-0482-0}{Lett.~Math.~Phys.~99,~231~(2012)}},
\texttt{\arxivref{1012.3992}{arxiv:1012.3992}}.
%%CITATION = 1012.3992;%%

\bibitem{Volin:2009uv}
D.~Volin,
\textit{``Minimal solution of the {$\AdS/\CFT$} crossing equation''},
\textsf{\doiref{10.1088/1751-8113/42/37/372001}{J.~Phys.~A42,~372001~(2009)}},
\texttt{\arxivref{0904.4929}{arxiv:0904.4929}}.
%%CITATION = ARXIV:0904.4929;%%

\bibitem{Arutyunov:2009kf}
G.~Arutyunov and S.~Frolov,
\textit{``The Dressing Factor and Crossing Equations''},
\textsf{\doiref{10.1088/1751-8113/42/42/425401}{J.~Phys.~A42,~425401~(2009)}},
\texttt{\arxivref{0904.4575}{arxiv:0904.4575}}.
%%CITATION = 0904.4575;%%

\bibitem{Fontanella:2019ury}
A.~Fontanella, O.~Ohlsson~Sax, B.~Stefa{\'n}ski,~jr. and A.~Torrielli,
\textit{``The Effectiveness of Relativistic Invariance in {$\AdS_3$}''},
\texttt{\arxivref{1905.00757}{arxiv:1905.00757}}.
%%CITATION = ARXIV:1905.00757;%%

\bibitem{Fontanella:2019baq}
A.~Fontanella and A.~Torrielli,
\textit{``Geometry of Massless Scattering in Integrable Superstring''},
\texttt{\arxivref{1903.10759}{arxiv:1903.10759}}.
%%CITATION = ARXIV:1903.10759;%%

\bibitem{Sax:2012jv}
O.~Ohlsson~Sax, B.~Stefa{\'n}ski,~jr. and A.~Torrielli,
\textit{``On the massless modes of the {$\AdS_3/\CFT_2$} integrable systems''},
\textsf{\doiref{10.1007/JHEP03(2013)109}{JHEP~1303,~109~(2013)}},
\texttt{\arxivref{1211.1952}{arxiv:1211.1952}}.
%%CITATION = ARXIV:1211.1952;%%

\bibitem{Bombardelli:2018jkj}
D.~Bombardelli, B.~Stefa{\'n}ski and A.~Torrielli,
\textit{``The low-energy limit of {$\AdS_3/\CFT_2$} and its {TBA}''},
\texttt{\arxivref{1807.07775}{arxiv:1807.07775}}.
%%CITATION = ARXIV:1807.07775;%%

\bibitem{Zamolodchikov:1978xm}
A.~B.~Zamolodchikov and A.~B.~Zamolodchikov,
\textit{``Factorized {S}-matrices in two dimensions as the exact solutions of
  certain relativistic quantum field models''},
\textsf{\doiref{10.1016/0003-4916(79)90391-9}{Annals~Phys.~120,~253~(1979)}}.
%%CITATION = APNYA,120,253;%%

\bibitem{OhlssonSax:2018hgc}
O.~Ohlsson~Sax and B.~Stefa{\'n}ski,~jr.,
\textit{``Closed Strings and Moduli in {$\AdS_3/\CFT_2$}''},
\textsf{\doiref{10.1007/JHEP05(2018)101}{JHEP~1805,~101~(2018)}},
\texttt{\arxivref{1804.02023}{arxiv:1804.02023}}.
%%CITATION = ARXIV:1804.02023;%%

\bibitem{Hoare:2013pma}
B.~Hoare and A.~A.~Tseytlin,
\textit{``On string theory on {$\AdS_3 \times \Sphere^3 \times \Torus^4$} with
  mixed 3-form flux: tree-level {S}-matrix''},
\textsf{\doiref{10.1016/j.nuclphysb.2013.05.005}{Nucl.~Phys.~B873,~682~(2013)}},
\texttt{\arxivref{1303.1037}{arxiv:1303.1037}}.
%%CITATION = ARXIV:1303.1037;%%

\bibitem{Hoare:2013ida}
B.~Hoare and A.~Tseytlin,
\textit{``Massive {S}-matrix of {$\AdS_3 \times \Sphere^3 \times \Torus^4$}
  superstring theory with mixed 3-form flux''},
\textsf{\doiref{10.1016/j.nuclphysb.2013.04.024}{Nucl.~Phys.~B873,~395~(2013)}},
\texttt{\arxivref{1304.4099}{arxiv:1304.4099}}.
%%CITATION = ARXIV:1304.4099;%%

\bibitem{Hoare:2013lja}
B.~Hoare, A.~Stepanchuk and A.~Tseytlin,
\textit{``Giant magnon solution and dispersion relation in string theory in
  {$\AdS_3 \times \Sphere^3 \times \Torus^4$} with mixed flux''},
\textsf{\doiref{10.1016/j.nuclphysb.2013.12.011}{Nucl.~Phys.~B879,~318~(2014)}},
\texttt{\arxivref{1311.1794}{arxiv:1311.1794}}.
%%CITATION = ARXIV:1311.1794;%%

\bibitem{OhlssonSax:2011ms}
O.~Ohlsson~Sax and B.~Stefa{\'n}ski,~jr.,
\textit{``Integrability, spin-chains and the {$\AdS_3/\CFT_2$}
  correspondence''},
\textsf{\doiref{10.1007/JHEP08(2011)029}{JHEP~1108,~029~(2011)}},
\texttt{\arxivref{1106.2558}{arxiv:1106.2558}}.
%%CITATION = 1106.2558;%%

\bibitem{Sundin:2012gc}
P.~Sundin and L.~Wulff,
\textit{``Classical integrability and quantum aspects of the {$\AdS_3 \times
  \Sphere^3 \times \Sphere^3 \times \Sphere^1$} superstring''},
\textsf{\doiref{10.1007/JHEP10(2012)109}{JHEP~1210,~109~(2012)}},
\texttt{\arxivref{1207.5531}{arxiv:1207.5531}}.
%%CITATION = ARXIV:1207.5531;%%

\bibitem{Borsato:2015mma}
R.~Borsato, O.~Ohlsson~Sax, A.~Sfondrini and B.~Stefa{\'n}ski,~jr.,
\textit{``The {$\AdS_3 \times \Sphere^3 \times \Sphere^3 \times \Sphere^1$}
  worldsheet {S} matrix''},
\textsf{\doiref{10.1088/1751-8113/48/41/415401}{J.~Phys.~A48,~415401~(2015)}},
\texttt{\arxivref{1506.00218}{arxiv:1506.00218}}.
%%CITATION = ARXIV:1506.00218;%%

\bibitem{Arutyunov:2007tc}
G.~Arutyunov and S.~Frolov,
\textit{``On String {S}-matrix, Bound States and {TBA}''},
\textsf{\doiref{10.1088/1126-6708/2007/12/024}{JHEP~0712,~024~(2007)}},
\texttt{\arxivref{0710.1568}{arxiv:0710.1568}}.
%%CITATION = ARXIV:0710.1568;%%

\end{thebibliography}

\end{document}